%% file: g1-j1226.tex
\documentclass[a4paper,fleqn,usenatbib]{mnras}

\usepackage{amsmath}
\usepackage{gensymb}
\usepackage{graphicx}
\graphicspath{{figs/}}
\usepackage{verbatim}
\usepackage{amssymb}
\usepackage{lineno}
\usepackage{wasysym}
\usepackage{amssymb}
\usepackage{xcolor}
\usepackage{soul}
\usepackage{amstext}
\usepackage{array}

\newcommand{\new}[1]{{#1}} % for 1st referee response
\newcommand{\neww}[1]{{#1}} % for 2nd referee response

\newcommand{\kms}{km~s$^{-1}$}

\newcommand{\mgii}{\ion{Mg}{ii}}
\newcommand{\Siii}{\ion{Si}{ii}}

\newcommand{\mgi}{\ion{Mg}{i}}
\newcommand{\hi}{\ion{H}{i}}
\newcommand{\feii}{\ion{Fe}{ii}}
\newcommand{\oii}{[\ion{O}{ii}]}

\newcommand{\ovi}{\ion{O}{vi}}

\newcommand{\sgas}{SGAS\,J1226+2152}

\newcommand{\zabs}{0.77138}  % updated to [OII] centroid
\newcommand{\zgal}{\zabs} % z_G1
\newcommand{\zarc}{2.9233} % z_arcs

\newcommand{\msun}{M$_{\astrosun}$}
\newcommand{\msunm}{{\rm M}_{\astrosun}}

\newcommand{\zsys}{$z_{\rm sys}$}

\newcommand{\ra}{12$^{\rm h}$26$^{\rm m}$51.31$^{\rm s}$}
\newcommand{\dec}{$+$21\degree52\arcmin17.22\arcsec}

\newcommand{\ragalpak}{12$^{\rm h}$26$^{\rm m}$51.46$^{\rm s}$}
\newcommand{\decgalpak}{$+$21\degree52\arcmin07.11\arcsec}

\newcommand{\inclstars}{53 \pm 10}  % b/a=0.80 +- 0.1  => incl = arcsen(b/a) 53 +/- 11, 9 
\newcommand{\pastars}{86 \pm 18}
\newcommand{\eradstars}{2.7}  % 0.37" => 0.37*7.41 kpc
\newcommand{\sersic}{0.42 \pm 0.16}

\newcommand{\LumB}{0.27 \pm 0.02~L^*_B}
\newcommand{\fluxoii}{$1.6 \pm 0.2 \times 10^{-17}$\,erg\,s$^{-1}$\,cm$^{-2}$}  % Giussepe # full [OII ]doublet
\newcommand{\sfroii}{$0.6 \pm 0.2$\,\msun\,yr$^{-1}$} %  SFR from OII doublet based on the K98 relation
\newcommand{\sfrsed}{$1.0 \pm 0.2$\,\msun\,yr$^{-1}$}  % SFR from Bagpipes
\newcommand{\ssfr}{$5-8 \times 10^{-11}$\,yr$^{-1}$} % sSFR NT
\newcommand{\massstars}{10.12 \pm 0.05}  % log scale Using Bagpipes estimate
\newcommand{\halomassstars}{11.7 \pm 0.2} % log scale e.g. from Moster 2010 % check this later
\newcommand{\rvir}{123 \pm 20}  % kpc
\newcommand{\vmaxhalo}{132 \pm 20} % kkm/s

\newcommand{\incloii}{48 \pm 1}
\newcommand{\paoii}{68 \pm 3} % includes the uncertainty in PSF modeling and oversampling
\newcommand{\troii}{0.1 \pm 0.1}  % in kpc?? check
\newcommand{\vmaxoii}{76 \pm 2}
\newcommand{\vdispoii}{33 \pm 2}
\newcommand{\ntot}{46} % arc spaxels with S/N
\newcommand{\nhit}{27} % arc detections
\newcommand{\nERDtwo}{17}  % 2sigma w/r to ERD model
\newcommand{\nERD}{23} % 3 sigma w/r to ERD model
\newcommand{\noutliers}{4}

\newcommand{\ew}{$W_0$}
\newcommand{\vel}{$v$}
\newcommand{\vdisp}{$\sigma$}
\newcommand{\ewm}{W_0}  % for equations
\newcommand{\velm}{v} % for equations
\newcommand{\vdispm}{\sigma}% for equations
\newcommand{\rvirm}{R_{\rm vir}}

\newcommand{\galpak}{{\sc Galpak}}
\newcommand{\megasaura}{{\sc Meg}a{\sc S}a{\sc ura}}
\newcommand{\pymuse}{{\sc PyMUSE}}
\newcommand{\marz}{{\sc MARZ}}
\newcommand{\sep}{{\sc SEP}}
\newcommand{\vpfit}{{\sc VPFit}}

\newcommand{\ppxf}{{\sc pPXF}}
\newcommand{\galfitm}{{\sc GalfitM}}
\newcommand{\lenstool}{{\tt{Lenstool}}}
\newcommand{\hst}{{\it HST}}
\newcommand{\HST}{{\it HST}}
\newcommand{\bagpipes}{{\sc Bagpipes}}

\title[CGM rotation \& outflows]{Telltale signs of metal recycling in the circumgalactic medium of a $z\sim 0.77$ galaxy}

\author[N. Tejos et al.]{N. Tejos,$^{1}$\thanks{E-mail: nicolas.tejos@pucv.cl}
S. L\'opez,$^{2}$
C. Ledoux,$^{3}$
A. Fern\'andez-Figueroa,$^{2}$
N. Rivas,$^{1}$\and
K. Sharon,$^{4}$
E. J. Johnston,$^{5,6}$
M. K. Florian,$^{7}$
G. D'Ago,$^{5}$
A. Katsianis,$^{8}$\and
F. Barrientos,$^{5}$
T. Berg,$^{3}$
F. Corro-Guerra,$^{1}$
M. Hamel,$^{2}$
C. Moya-Sierralta,$^{5}$\and
S. Poudel,$^{1}$
J. R. Rigby,$^{7}$ and
M. Solimano$^{5,6}$\and
\\
$^{1}$ Instituto de F\'isica, Pontificia Universidad Cat\'olica de Valpara\'iso, Casilla 4059, Valpara\'iso, Chile\\
$^{2}$ Departamento de Astronom\'ia, Universidad de Chile, Casilla 36-D, Santiago, Chile\\
$^{3}$ European Southern Observatory, Alonso de C\'ordova 3107, Vitacura, Casilla 19001, Santiago, Chile\\
$^{4}$ Department of Astronomy, University of Michigan, 1085 South University Avenue, Ann Arbor, MI 48109, USA\\
$^{5}$ Instituto de Astrof\'isica, Pontificia Universidad Cat\'olica de Chile, Av.~Vicu\~na Mackenna 4860, 7820436 Macul,\\ \ \ Santiago, Chile \\
$^{6}$ N\'ucleo de Astronom\'ia de la Facultad de Ingenier\'ia y Ciencias, Universidad Diego Portales, Av. Ej\'ercito Libertador 441,\\ \ \ Santiago, Chile\\
$^{7}$ Observational Cosmology Lab, NASA Goddard Space Flight Center, Greenbelt, MD~20771, USA\\
$^{8}$ Tsung-Dao Lee Institute, Shanghai Jiao Tong University, Shanghai 200240, China\\
}

\begin{document}
\label{firstpage}
\pagerange{\pageref{firstpage}--\pageref{lastpage}}
\maketitle

\begin{abstract}
\input{abstract_v2}
\end{abstract}


\section{Introduction} 
\label{introduction}

% - importance of CGM
The diffuse circumgalactic medium (CGM) is currently recognized as a prime environment to understand galaxy evolution (see \citealt{Tumlinson2017}\neww{; \citealt{Peroux2020} for recent reviews}). 

% challenges to observe the CGM
% - 1d versus tomographic
Its %general gas density 
average absorbing properties have been established with great significance from traditional quasar absorption-line surveys of galaxy-quasar pairs \citep[e.g.][]{Bahcall1969, Bergeron1986, Bergeron1991, Lanzetta1995, Chen1998, Churchill2000, Werk2014, Prochaska2017, Chen2017, Dutta2020}, but these studies lack the spatial information on individual galaxy haloes, thus hampering a link between those \new{circumgalactic} media and their associated galaxies. For instance, resolving the CGM gas kinematics to probe the baryon cycle (inflows, rotation, outflows) becomes particularly challenging, \neww{because} of the geometrical degeneracy of having rest-frame
velocity offsets for only one single impact parameter to a galaxy without knowing whether the absorbing gas is foreground or background to its host.
%, it is not possible to establish whether the gas is inflowing, outflowing, rotating, etc. 
Despite this inherent difficulty, some efforts exist, including the use of multiple single quasar sightlines per galaxy \citep[e.g.][]{Nielsen2013, Chen2014, Bowen2016, Zahedy2016, Rubin2018c, Kulkarni2019, Lehner2020}, the use of galaxies as extended background sources \citep[e.g.][]{Steidel2010, Diamond-Stanic2016, Rubin2018b, Peroux2018, Dupuis2021}, and/or those that rely on priors like metallicity of the absorbing clouds \citep[e.g.][]{Ribaudo2011}, or the relative geometry of the galaxy observed through stellar emission with respect to the quasar sightline \citep[e.g.][]{Bouche2012, Kacprzak2012, Ho2017, Martin2019, Ho2020, Zabl2019}. However, even if these priors are correct, having only few discrete points per galaxy halo precludes  significant progress on resolving the complex CGM structure and its kinematics.

A more certain case is when the absorbing gas is observed ``down-the-barrel''---i.e. using the targeted galaxy itself as background source---for which we know that the gas must be foreground and thus signatures for inflows and/or outflows can be disentangled \citep[][]{Rubin2012, Kornei2012, Talia2012,  Martin2012, Rubin2014, Kacprzak2014, Heckman2015}. However, in this latter case, what remains unknown is the location of the absorbing gas, which would most likely be confined to the denser ISM or halo/disc interface \citep[e.g.][]{Wakker1997, Bregman1980, Putman2012}; still, exceptional cases have been reported towards star-forming galaxies revealing P-Cygni emission/absorption profiles even on (inner) CGM scales \citep[e.g.][]{Rubin2011, Burchett2021}.

% tomographic, clumpy and anisotropic + hints of rotation
\neww{To help  circumvent these issues, in} the past few years we have developed a complementary approach using giant bright gravitational
arcs as background \neww{sources} \citep[][hereafter referred to as Paper~I and Paper~II, respectively]{Lopez2018, Lopez2020}. The great advantage of this method is that it provides a tomographic view of the \neww{intervening gas,} i.e., it yields an ensemble of contiguous impact parameters, velocity offsets and velocity dispersion for {\it individual} galaxy halos. %, and that is independent of the CGM kinematics. 
These configurations
%Such rich amount of information not only allow us to directly assess the clumpiness and anisotropy of the CGM in individual galaxy halos (greatly complementary to the results from quasar absorption lines), but it 
break geometrical degeneracies and \neww{have been} used to resolve the gas kinematics, assess anisotropy, and determine clumpiness and covering fraction of the CGM of individual galaxies. Ultimately, gravitational-arc tomography 
provides \new{complementary} observational evidence to link halo gas and host galaxy properties. 

% This paper
In this third paper we focus on the kinematics of the
metal-enriched cool-warm ($T\sim 10^{3-4}$\,K) CGM of an isolated star-forming galaxy at $z=\zabs$. The galaxy lies close in
projection to the gravitational arcs formed by the \sgas\ system at $z=2.9233$
\citep{koester2010, Rigby2018}\new{, on which we apply the arc-tomography technique \citep[see also][]{Mortensen2021}}. The unique geometrical configuration of the arc segments around the
galaxy allows us to probe the CGM at impact parameters of $D\approx 10-30$\,kpc (mostly along the minor-axis) and $D\approx 60$\,kpc (along the major-axis), whilst covering a wide range of azimuthal angles.

%structure
The paper is structured as follows. In \S~\ref{sec:obs}, we present the target selection
and \neww{our} observations. We describe the field-of-view and give details on the
adopted lens model in \S~\ref{sec:field}. In \S~\ref{sec:emission}, we describe the
properties of the intervening absorbing galaxy and in \S~\ref{sec:abs} we present the
methodology for analysing the absorption lines. In \S~\ref{sec:results}, we present our
main results which we discuss in \S~\ref{sec:discussion}. A summary and conclusion is presented 
in \S~\ref{sec:summary}. In \neww{the Appendix}, we present complementary \neww{analysis}.
Throughout, we assume a flat $\Lambda$CDM cosmology with $\Omega_{\rm m}=0.3$,
$\Omega_{\Lambda}=0.7$, and $H_0=70$\,km\,s$^{-1}$\,Mpc$^{-1}$.

\input{fig_muse}

\section{Observations}\label{sec:obs}

\subsection{Target selection}

Our main target is the galaxy that lies between the arcs of \sgas\ \citep{koester2010}, hereafter referred to as G1 (see Fig.~\ref{fig:fov}). The arcs and G1 were selected from the \megasaura\ survey \citep{Rigby2018},  which followed-up a
sample of bright gravitational arcs with the Magellan/MagE
spectrograph \citep{Marshall2008}.  \megasaura\ confirmed the redshifts of G1 at $z \approx 0.77$ and the Northern and
Southern arcs (hereafter referred to as arc-N and arc-S respectively) both at $z\approx 2.92$. 

The MagE spectrum from \megasaura\ revealed the presence of a strong \mgii\ absorber at the
redshift of G1 in arc-N, albeit at low spatial resolution (slit width of 2\arcsec\ over
multiple position angles; see fig.~1 from \citealp{Rigby2018}). The confirmed \mgii\ absorption plus the
geometrical configuration of extended arcs on opposite sides of G1 (i.e., somewhat
aligned with its minor axis) made this galaxy a primary target for our \neww{arc-tomography program}. 

\subsection{VLT/MUSE$+$AO}

We observed the field around G1 with the Multi-Unit Spectroscopic Explorer \citep[MUSE, ][]{Bacon_2010} at the Very Large Telescope (VLT) between April 2018 and January 2019, as part of program  101.A-0364 (PI: Lopez). The observations were carried out using the Wide-Field Adaptive-Optics \new{with the extended wavelength} mode (\new{WFM-AO-E}), providing a field of view (FoV) of $1\arcmin\times1$\arcmin\ with a pixelscale of $0.2$\arcsec, and \new{a wavelength range of $4600-9350$\,\AA\ with} a spectral resolving power of \textit{R}$\sim$2100 at 5000\,\AA\ (the wavelength of the \mgii\ absorption at $z\sim 0.77$). A set of $20 \times 640$\,s individual exposures were taken, applying a small spatial dithering and a position angle rotation of $90$\degree relative to the previous exposure in order to reduce the appearance of the image slicers and channels in the final stacked datacube. 

The data was reduced using the MUSE pipeline \citep{Weilbacher_2012} within the ESO Recipe Execution Tool (EsoRex) environment \citep{ESOREX}, using standard procedures and calibrations. The wavelength solution was calibrated to air \neww{(Table~\ref{table:wv_air} in the Appendix provides the rest-frame wavelengths for Mg~{\sc ii} and [O~{\sc ii}] used in the subsequent analysis)}.
The residual sky contamination from the standard reduction was measured and cleaned using the Zurich Atmosphere Purge code \citep[ZAP,][]{Soto_2016}. We finally aligned the WCS of the datacube to match that of the HST imaging (see below) using a bright star in the FoV as reference. \new{The total integration time of the datacube is $3.6$\,h and the effective FWHM PSF is 0.7\arcsec, as measured from the $V$-band image created from the datacube by applying a Gaussian fit to a star within the field.}

\subsection{\HST/WFC3 and \HST/ACS}\label{sec:hst}

Hubble Space Telescope (\hst) data were taken in five different bands from three different GO programs. F160W and F110W observations were carried out for 1312\,s and 1212\,s, respectively, with the WFC3 IR channel for the program GO~15378 (PI: Bayliss). Additional imaging in the WFC3 F110W band was carried out as part of a SNAP program, GO~12116 (PI: Ebeling), for 706\,s, offset from and not including the main arc or absorber. This program also acquired 706\,s of imaging in the F140W band with a similar offset pointing, as well as 1200\,s of imaging in the ACS F606W band and 1440\,s in the ACS F814W band that did include the arc segments and absorber.  The latter two bands (F606W and F814W) were also observed in two visits each (for 2040\,s per visit, i.e. a total of 4080\,s in each band) by the program GO 12368 (PI: Morris).   This program also observed an offset position of the larger cluster with the WFC3 UVIS F814W and F606W filters for 8033\,s and 5888\,s respectively.  The offset fields from programs GO~12368 and GO~12166 were used in modeling the lensing effects of other pieces of the cluster, including the secondary halo to the South described in \S~\ref{sec:lens}.

\neww{The \HST\ }data were reduced using the {\sc DrizzlePac} package.\footnote{\url{https://www.stsci.edu/scientific-community/software/drizzlepac}}  Images taken with the same instrument and filter but at different epochs were each drizzled, using {\tt astrodrizzle}, onto $0.03$ arcsec/pixel grids, one for each epoch.  These drizzled products were then aligned with each other, using the {\tt tweakreg} routine.  The resulting WCS solutions \new{are within $\approx 0.1$\arcsec\ to that of Gaia DR2 \citep{GaiaDR2, GaiaDR2_astro}}, and were propagated back to the individual flatfield-calibrated files using the {\tt tweakback} routine. Those individual flatfield-calibrated frames were then all drizzled together onto a common pixel grid with a $0.03$ arcsec/pixel scale using {\tt astrodrizzle} with a Gaussian kernel, and a final drop size of $0.8$.  Data from other instrument and filter combinations were drizzled using the same {\tt astrodrizzle} parameters before being aligned to a common reference grid defined by the ACS~F814W band data from the program GO 12368 because it provided the best alignment with the offset F814W~WFC3 data which were taken at the same time for the same program.  Ultimately, this process resulted in a single epoch of observations of the arc in both the WFC3 F160W and F110W filters (from GO~15378), a stack of two epochs of imaging in the ACS F814W and F606W filters from the GO programs 12166 and 12368, and offset observations of other areas of the cluster in the WFC3 UVIS channel's F606W and F814W bands and the WFC3 IR channel's F110W and F140W bands. 

\subsection{Magellan/MagE}\label{sec:mage_main}

\neww{We observed \sgas\ with the MagE Echellette spectrograph~\citep{Marshall2008} on the $6.5$\,m Magellan telescopes on the nights of March 10 and 11, 2019, under clear skies and variable seeing conditions ($0.8$\arcsec--$1.4$\arcsec). We used two $1$\arcsec\ wide slits, which we refer to as the Eastern (``E'') and Western (``W'') slits (see Fig.~\ref{fig:mage_slits} in the Appendix). We integrated on target for a total of $4.8$\,h and $5.0$\,h on ``E'' and ``W'', respectively. 

Data reduction proceeded as in \citet{Lopez2020}, with a wavelength solution calibrated to air \new{using the `echelle' package of ESO/MIDAS \citep{midas_echelle}}. The resolving power of the final spectra is $R=4\,500$ with a dispersion of $0.37187$\,\AA\,pixel$^{-1}$. The reduced and spatially-binned
spectra provide pseudo-spaxels of $0.9\arcsec\times1.0\arcsec$ in size at $11$ independent positions along both ``E'' and ``W'' (Fig.~\ref{fig:mage_slits}). The spectral resolution of MagE is twice that of MUSE, allowing us to better resolve the Mg~{\sc ii} absorption profiles at a comparable spatial sampling.
}

\section{Field and lens model}
\label{sec:field}

\subsection{Background regions}\label{sec:arcs}

For our main analysis we define $4$ regions with background signal for probing \mgii\ at $z=\zabs$: three arc segments 
\neww{providing transverse probes} at the North, South and East of G1 (hereafter referred to as `arc-N', `arc-S', `arc-E', respectively), and G1 itself, \neww{providing} `down-the-barrel' \neww{probes} (hereafter referred to as `DtB'). These regions, along with our definition of \mgii\ \neww{binned} spaxels (\S~\ref{sec:binning}), \neww{have been color coded} in \neww{Figs.~\ref{fig:fov},~\ref{fig:antonia}~and~\ref{fig:kinematics}}

\subsection{Galaxies at $z\approx 0.77$}

We performed a blind galaxy survey within the $1\arcmin \times 1\arcmin$ MUSE FoV, in a similar fashion as in Paper~I. Continuum sources were searched for using the Source Extractor and Photometry software \citep[\sep;][]{Barbary2018, Bertin1996} on the deeper HST F814W image, and their 1-d spectra were optimally extracted with \pymuse\ \citep{Pessa2020} from within these detected regions in the MUSE data. We then used the template-matching algorithm \marz\ \citep{Hinton2016} to obtain their spectroscopic redshifts. \neww{We also searched for emission-line only sources by performing a pseudo-narrow band image centred at the [O~{\sc ii}] emission line at the redshift of G1.}

We found \neww{only} two galaxies at $z\approx 0.77$, including G1 itself; the second galaxy (referred to as G2) appears at $\approx 38$\arcsec\ from G1 in the image plane, i.e. right at the Northern edge of the MUSE FoV (not shown \neww{in Fig.~\ref{fig:fov}}). This angular separation corresponds to a physical proper transverse distance of $225$\,kpc (delensed; see \S~\ref{sec:lens}).

\neww{Our galaxy survey is} $\sim 75\%$ complete for galaxies brighter than $m_r\sim 23$\,mag. We have also placed an emission line flux limit of $1.7 \times 10^{-18}$\,erg\,s$^{-1}$\,cm$^{-2}$ at $\lambda_{\rm obs}=6630$\,\AA, the [O~{\sc ii}]\ wavelength at G1's redshift. As G1 is located near the centre of the MUSE FoV, we conclude that the galaxy is relatively isolated within $\approx 220$\,kpc. However, one may see from the \oii\ emission \neww{(see \S~\ref{sec:emission})} and from the \HST\ imaging that G1 may be in the process of accreting a small satellite \neww{(referred to as ``G1-s''; see Figs.~\ref{fig:fov}~and~\ref{fig:oii_2D}).}

\subsection{Lens model}\label{sec:lens}

The lensing analysis is based on archival \HST\ imaging data (\S~\ref{sec:hst}). 
The lens plane was modeled using the \lenstool\ software \citep{Jullo2007}. \lenstool\ uses a parametric approach, in which each mass component is represented by a parametric function. %, whose parameters are allowed to vary. 
The software explores the parameter space and establishes the best-fit parameter set using a Markov Chain Monte Carlo (MCMC) algorithm. 

The cluster lens, \sgas\ \new{lies at $z=0.43$ \citep{Bayliss2011}, and is} modeled with several cluster-scale halos, each assumed to be a pseudo-isothermal ellipsoidal mass distribution (PIEMD) with six free parameters: $x$, $y$, ellipticity $e$, position angle, core radius $r_{core}$, truncation radius $r_{cut}$ and a normalization $\sigma_0$. \new{Galactic}-scale potentials are also parameterized as PIEMD, whose positional parameters remain fixed during the minimization process, and slope parameters are scaled to the observed luminosity. The cluster members were selected according to their F606W-F814W colors in a color-magnitude diagram, following \cite{Gladders2000}. 

The photometry, spectroscopy, and lensing analysis of \sgas\ indicates that the foreground lens is quite complex. In addition to the massive halo that dominates the lensing potential near our main lensed targets ($\approx 10$\arcsec\ to the North-East; see Fig.~\ref{fig:fov}), we observe a massive, strong-lensing halo 153\arcsec\ to the South (not shown in the figure), at the same redshift \citep{Bayliss2011}.

In this work, we treat the southern massive halo as contributing shear to the lensing potential, and model it as a spherical PIEMD at (R.A., Decl.) = (12$^{\rm h}$26$^{\rm m}$51.11$^{\rm s}$, $+$21\degree49\arcmin52.3\arcsec)~J2000. 

Two additional galaxies are included as individual halos in the model. A perturber next to the East part of arc-N (at $z=2.9233$), and \neww{G1 itself}. We treat G1 as residing at the same plane of the cluster. We test models with and without this galaxy added as a lens component, and conclude that the results of this paper are not sensitive to this modeling choice.

We use as lensing constraints the three images of one of the lensed galaxies at $z=2.9233$ \citep{koester2010} (i.e., 2 images in arc-N plus the one associated to arc-E). The \hst\ imaging allows identification of individual emission knots in each of the images, to be used as constraints.

We \neww{computed} the source location of both the absorber G1 and the $z=\zarc$ source at the plane of the absorber at $z=\zabs$ \neww{(i.e. the absorber-plane)}, by ray-tracing through the lens equation, $\vec\beta=\vec\theta-d_{ls}/d_s\vec\alpha(\vec\theta)$, where $\vec\beta$ is the source position, $\vec\theta$ is the image position, $\vec\alpha$ is the deflection field from the model, and $d_{ls}$ and $d_s$ are the angular diameter distances between the lens and the source, and the observer and the source, respectively. The result of this ray-tracing process is shown Fig.~\ref{fig:fov} (right panels). Because $d_{ls}/d_s$ for the absorber-plane is much smaller than $d_{ls}/d_s$ for $z_{\rm source}=\zarc$, the deflection values are relatively small, as can be seen in Fig.~\ref{fig:fov}.

The resulting ray-traced field is used to calculate the impact parameters between G1 and the positions along the arc segments in the absorber-plane.

Uncertainties on all the lensing-derived quantities are obtained from a suite of lens model realizations, derived from steps in the MCMC sampling of the parameter space.

\input{galaxy_g1}

\input{fig_mgii_spectra}

\section{Absorption-line analysis}\label{sec:abs}

In this section, we provide a detailed description of our \mgii\ absorption-line analysis based on
MUSE data. \neww{For the sake of clarity, the analysis based on MagE data is presented in Appendix~\ref{sec:mage}.}

\subsection{Definition of spaxels for Mg\,{\sc ii} analysis}\label{sec:binning}
% mention why and how we define spaxels by binning 4x4 MUSE native

For the \mgii\ absorption analysis, we define squared apertures of $0.8\arcsec \times 0.8\arcsec$ in size in the image plane by 
re-binning the cube signal  optimally in $4 \times 4$ MUSE \neww{native}  spaxels. This has two main advantages compared to just using native (or oversampled) spaxels (Paper~I): (i) it increases the signal-to-noise of the rebinned spectra, and (ii) it significantly reduces the spatial \new{correlation} of contiguous spaxels. Indeed, given that the effective MUSE PSF FWHM=0.7\arcsec\ in the image plane, we estimate a \new{typical} of $\sim 5$\,per cent level of cross-talk between spaxels due to the ``wings'' of the Moffat function~\citep[for any reasonable value of its $\beta$ index; e.g.,][]{Trujillo2001};\footnote{\new{This is estimated by integrating the Moffat function described in eq.~1 of \citet{Trujillo2001} between $x_{\rm min}=0$ and $x_{\rm max}=0.4\arcsec/0.35\arcsec \approx 1.14$, with $x$ in units of the FWHM, and comparing it with the integral over the full domain.}} hence, \new{assuming sources near the center of the binned spaxels,} these apertures \new{should} result in mostly {\it independent} observational data points without significant loss of information.

\subsection{Selection of spaxels for Mg\,{\sc ii} analysis}

To perform a robust \neww{and} unbiased analysis, we pre-select spaxels based solely on signal-to-noise ($S/N$). The $S/N$ is measured using the RMS of the continuum within a spectral window close to \mgii\ and free of absorption features. A total of $\ntot$ independent spaxels are selected in this fashion to have $S/N>3$ \new{per $1.25$\,\AA\ pixel}: $16$ in arc-N, $21$ in arc-S, $5$ in arc-E, and $4$ in G1 DtB. The positions of these spaxels with respect to the arcs and G1 are presented in Fig.~\ref{fig:fov}.

We recall that these apertures have areas of a few kpc$^2$ in the absorber-plane; therefore, the observed absorption profile is the result of a complex convolution of \neww{the} light-weighted \neww{signal} by an extended source. \neww{As a consequence}, comparisons with the quasar statistics of \mgii\ absorption systems must be done with care, as we discuss below.

\subsection{Line-profile fitting}

Given the limited spectral resolution of our MUSE data, the absorption profiles are dominated (albeit not determined) by the instrument line spread function (\neww{LSF;} FWHM$\approx 2.7$\,\AA\ at \mgii). We \neww{derived} rest-frame equivalent width (\ew), centroid velocity with respect to \zsys\ (\vel) and velocity dispersion ($\sigma$)\footnote{\new{Velocity dispersions are obtained after deconvolving with the instrumental LSF.}} by attempting a fit to the \mgii\ doublet absorption {\it automatically} in all $46$ spaxels with inverted double-Gaussian profiles. \new{We estimated the local continuum by fitting a straight line to featureless regions on both sides of the expected Mg~{\sc ii}\ wavelengths, and normalized the spectra before the double-Gaussian fitting.} Wavelength ratio is tied in each fit, but not the line widths nor the doublet ratios. At MUSE resolution, \mgii\ is partially blended with \Siii~$\lambda 1260$ absorption at the source galaxy, thus we include a third Gaussian with free-varying redshift in all arc spaxels. In the \neww{DtB} region the \Siii~$\lambda 1260$ is not present and thus we do not include this third Gaussian in those fits. In some regions of arc-S, a Si~{\sc ii}*~$\lambda 1264$ emission line at the source galaxy is present between the two \mgii\ transitions of the doublet, but here we do not model it. \new{This choice may impact the fits, but we expect this to be a minor effect given that our model includes both members of the Mg~{\sc ii} doublet simultaneously.}

\input{fig_mgii_2D}

Our automatic fitting delivers $29$ fits deemed ``successful'' according to significance $>1.5$ in both \mgii\ transitions: $27$ towards arc segments, and $2$ DtB. For the rest of the spaxels, \ew\ $2\sigma$ upper-limits are calculated and reported. We visually inspected all the fits. They are shown in red in Fig.~\ref{fig:antonia}; the measurements, errors and upper limits are given in Table~\ref{tab:mgii}. Despite the low spectral resolution and S/N, our fits are in general robust in terms of \ew\ and \vel, but less so in \vdisp. Indeed, measured velocity dispersions must be considered with caution, as they are subject to systematics and are correlated with \ew\ by construction. 

Finally, delensed maps are constructed based on each spaxel position and impact parameters\neww{, $D$,} are calculated for each spaxel in the absorber-plane.

\input{table_mgii_nt}

\section{Results}\label{sec:results}

The geometrical configuration in \sgas\  
%of the arcs 
allows us to probe the CGM on opposite sides of G1.
%, specially along its minor axis. 
The first and second panels of Fig.~\ref{fig:mgii_2D} present 
%two-dimensional 
absorber-plane 
maps of continuum signal-to-noise and \ew\ measurements, respectively. We observe significant \mgii\ absorption around G1 at impact parameters within $\sim20$\,kpc, especially along the minor-axis (i.e. \neww{roughly the} North and South directions). In contrast, the fraction of detections decreases at larger impact parameters ($D$) and/or azimuthal angles ($\alpha$)\footnote{Defined as the angle between the semi-major axis of \new{the receding side of} the galaxy to the centre of the spaxel.} closer \new{to $180$\degree,} as seen from spaxels located towards the East of arc-S. A more detailed analysis of the \ew-$D$ and \ew-$\alpha$ relations will be presented \neww{elsewhere}, and in this paper we focus on the main kinematic signatures of the \mgii\ absorbing gas.

\subsection{Kinematics of the absorbing gas}\label{sec:kinematics}

The two rightmost panels of Fig.~\ref{fig:mgii_2D} show the absorber-plane velocity (\vel) and velocity dispersion (\vdisp) maps of the \mgii\ absorption, respectively. We observe that with respect to \zsys$=$\zabs, absorption towards arc-N show negative velocity offsets, with a mean value of $\bar{\velm} \approx -31 \pm 25$\,\kms\ (median of $-25$\,\kms) while absorption towards arc-S span both positive and negative velocity offsets, ranging from $\velm \approx +120$\,\kms\ on the West to $\velm \approx -100$\,\kms\ towards the East, with a mean value of $\bar{\velm} \approx +19 \pm 51$\,\kms (median of $+10$\,\kms). Regarding velocity dispersion, we find larger values towards the arc-N compared to towards arc-S (a factor of $2\times$ larger approximately), with mean values of $\bar{\vdispm} \approx 54 \pm 31$\,\kms\ (median $59$\,\kms)  and $\bar{\vdispm} \approx 29 \pm 31$\,\kms\ (median $29$\,\kms), respectively; but we caution the reader that the MUSE LSF dominates the profiles. \neww{Our} typical values for \vdisp\ are consistent with those reported by \citet{Mortensen2021} based on independent observations of the same arc using KCWI with a similar LSF \new{(see \S~\ref{sec:mortensen} for further comparisons)}.

\new{Given that arc-N is mostly aligned with the semi-minor axis of G1, one may expect a galactic outflow to \neww{affect} its kinematic trends. However, we find i) small typical velocity centroids ($|\velm| \sim 20-25$\,\kms) for G1's \neww{low inclination}, and ii) small velocity centroids dispersion over tens of kpc (i.e. across different spaxels).} In the following, we explore whether a simple extended rotating disc (ERD) model can explain the observed \mgii\ kinematic trend\new{, but in \S~\ref{sec:vdisp} and \S~\ref{sec:inflow_outflow} we come back to the galactic outflow possibility.}

\subsection{Extended rotating-disc (ERD) model}\label{sec:ERD}

We start from the \new{3-dimensional rotating-disc model obtained \neww{for [O~{\sc ii}] (\S~\ref{sec:galpak}) and} produce an infinitely thin disc model with the same main relevant parameters (namely, PA$_{\rm gas}$, $i_{\rm gas}$, $r_{\rm t}$, and $v_{\rm max}$; see Table~\ref{tab:G1}). We then  extrapolate this simplified model} to much larger distances \neww{allowing} us to cover the arcs in projection. Because we are not actually {\it fitting} the absorption profiles with the disc model, we simply take the velocities from the model at mid-plane to compare them to the observed velocity centroids (\vel). With this (over)simplification we \new{are assuming that the velocity-weighted optical depth of Mg~{\sc ii} absorption is dominated by gas located close to the \neww{putative} disc mid-plane (i.e., that $H_{\rm eff} \ll 1$; see~\citealp{Ho2017}).}

\input{figs_kinematics}

In the bottom-right panel of Fig.~\ref{fig:oii_2D} we show the centroid velocities of the starting (inner) velocity model 
%in relation to G1 
(in the absorber-plane and using native MUSE spaxels). These are to be compared with the actual {observed} velocity of \oii\ emission,
%. We observe a good qualitative agreement, despite the differences between our ERD model and that obtained from \galpak; in particular, 
although we emphasize that this direct comparison is an oversimplification because of the spatial correlations introduced by the resampling and coarse binning of the input datacube in the morpho-kinematic analysis \neww{(see \S~\ref{sec:galpak}). }

\subsection{Comparison of Mg\,{\sc ii} kinematics to the ERD model}\label{sec:ERDcomp}

We now proceed to compare the ERD model with the actual absorbing \mgii\ kinematics. In the left panel of Fig.~\ref{fig:kinematics}, we show the velocity field of our simple ERD model around G1. In order to account for the fact that the observed \mgii\ signal comes from the integration over extended areas, we consider a velocity range given by the maximum and minimum velocity allowed by the ERD model in a given spaxel; the errors in the main parameters of the ERD model \new{(obtained from the \galpak\ analysis)} are contained within the aforementioned allowed velocity ranges in a given spaxel. In the right panel of Fig.~\ref{fig:kinematics}, we show the velocity centroids of \mgii\ absorption (\vel) as a function of impact parameter ($D$), together with the velocity range allowed by the model for each spaxel; \oii\ emission measurements are also shown for reference. Positive impact parameters are arbitrarily defined to be on the receding side of the velocity field, and negative values otherwise. We emphasize that this plot {is not} a rotation curve, as our spaxels are located in a wide range of azimuthal angles with respect to the semi-major axis of G1.

We observe that the vast majority of the velocity measurements lie on the `correct' side of the model velocity field (either positive or negative, as defined above) within uncertainty. Moreover, out of a total of $\nhit$  measurements, $\nERDtwo$ ($\nERD$) lie within $2\sigma$($3\sigma$) from the model expectation, \neww{corresponding to a large} ``kinematic matching fraction'' of $\sim 60\%$($85\%$). \neww{This strongly supports} the scenario of co-rotating \mgii.

We have run tests varying the main parameters of the ERD model (PA$_{\rm gas}$, $i_{\rm gas}$, $v_{\rm max}$) within $\pm 10\%$ and we still find consistency with the data  at a similar level. This illustrates that although a good match exists, our current dataset (and modeling)
does not constrain the geometry of the putative extended disc to better than $10\%$. \neww{This} is not surprising, as most of the spaxels are probing directions aligned with the semi-minor axis at large impact parameters where a kinematical model is less sensible. However, this level of consistency between the simple ERD model and the observations does provide observational evidence for the existence of an extended rotating disc traced by \mgii\ in absorption that co-rotates with the inner disc traced by \oii\ in emission. We discuss this result in more detail in \S~\ref{sec:corotating}.

\input{fig_mgii_mage_main}

% \subsection{Down-the-barrel Mg~{\sc ii} absorption}
\subsection{Velocity dispersion and kinematic outliers}\label{sec:vdisp}

\new{Although the ERD model seems a viable scenario for explaining a large fraction of the \neww{observed centroid velocities,} it cannot account for the velocity dispersion (\vdisp) signatures \neww{(by construction)}. In general, we} observe large \vdisp\ on both arcs, typically ranging from $\vdispm \approx 30-100$\,\kms\ (see above). These values are comparable, or even larger than what is expected from the rotation alone, thus indicating the presence of unresolved non-rotational components. Indeed, we find typical $v_{\rm rot}/\vdispm \sim 0.5-2$ indicative of a turbulent medium \citep[e.g.][and references therein; see \S~\ref{sec:inflow_outflow} for further discussion on this]{Glazebrook2013}. 

%DtB absorption link to outflow
% \SL{separate sub-section?}

A particularly interesting region in the context of velocity dispersion, is that observed ``down-the-barrel'' (DtB) to G1. The DtB's \ew\ and \vdisp\ values are the largest, reaching $\sim 2-3$\,\AA\ and $160-170$\,\kms\ respectively (see Figs.~\ref{fig:antonia},~\ref{fig:mgii_2D},~\ref{fig:mage_main} and Table~\ref{tab:mgii}). Clearly, the large values in \ew\ are driven by the velocity dispersion; however, from the limited $S/N$ and spectral resolution of MUSE we cannot resolve the individual components in these DtB spaxels.

\neww{Our MagE data provide} $2 \times$ better spectral resolution than that of the MUSE data. In Appendix~\ref{sec:mage}, we present the details of \neww{the} analysis based on Voigt profile fitting, where only arc-N and G1 provided enough $S/N$ for meaningful fits. From these data, we can resolve at least two kinematical components in arc-N and G1 DtB as shown in Fig.~\ref{fig:mage_main} (but see Fig.~\ref{fig:magefits} for further absorption profiles on different positions in arc-N). For both, arc-N and DtB regions, we observe that one \mgii\ absorption component lies at velocity consistent with that expected from the ERD model at their respective position (i.e. at $\velm \sim -20 - 0$\,\kms), while the second component is significantly blue-shifted: at $\velm \sim -380$\,\kms\ for G1 DtB, and at $\velm \sim -140$\,\kms\ for arc-N.

We observe that G1 DtB is very complex kinematically, and that even at the resolution of MagE, we can only partially resolve it. On the other hand, the absorption profile seen in arc-N is in general less broad \neww{although} we emphasize that the second bluer \new{component} is observed all over arc-N, and at similar large absolute velocities ($\velm \approx -140$\,\kms; see Table~\ref{table:mgii_mage} and Fig.~\ref{fig:magefits}). At the resolution of MUSE, the second blue-shifted components would get diluted producing large velocity dispersions for a single absorption \new{component} fit.

% Kinematic outliers
\neww{Coming back to the MUSE data and} considering the \mgii\ absorption in the arc segments, we now focus on those that have velocity centroids (\vel) inconsistent with the ERD model at the $3\sigma$ level. These correspond to $2$ spaxels in arc-N and $2$ spaxels in arc-S, for a total of $\noutliers$ out of $\nhit$ spaxels with detections (see \S~\ref{sec:ERDcomp}). We deem these points as ``kinematic outliers'' and are marked with a `n' in the last column of Table~\ref{tab:mgii} and a \neww{cross} symbol in Fig.~\ref{fig:kinematics}. These datapoints are of particular interest because they can represent direct signatures of inflows or outflows in the CGM of G1. In \S~\ref{sec:inflow_outflow} we discuss this possibility in the context of a simple galactic outflow model that could physically link the blue component of G1 DtB with the blue components seen in arc-N.

\section{Discussion}\label{sec:discussion}

\subsection{Is the bulk of Mg~{\sc ii} co-rotating?}\label{sec:corotating}

In \S~\ref{sec:ERDcomp} we have shown that a simple extended rotating-disc (ERD) model can explain most of the \mgii\ \neww{line centroids},  
thus providing evidence for the presence of an extended co-rotating disc in the CGM of G1. Considering the projected distances (mostly along the minor-axis) and the inclination angle of the disc ($i\sim 48$\degree), we are probing co-rotating \mgii\ at radial distances out to $r \approx D / \cos(i) \approx 42$\,kpc along the disc (assuming $D\sim 23$\,kpc and $\alpha \sim 90\degree$), corresponding to a physical extent of $\approx 15 R_e$ and about $\sim 0.4\,\rvirm$.

\neww{Although the evidence for extended rotating discs is not new, here we emphasize that our approach is \new{complementary} to previous experiments. These}
large discs are common at low-$z$, seen typically from \hi~21-cm emission \new{\citep[e.g.][and references therein]{Sancisi2008},} and some evidence of extended co-rotating CGM from \hi\ absorption at $z <0.03$ with  distances $\lesssim \rvirm$ has been recently \neww{presented} \citep{French2020}. At $z\sim0.2-0.4$ some hints from extrapolating a rotation curve to single impact parameters \neww{have been} given by single or pairs of quasar per galaxy \citep[][]{Martin2019, Ho2017, Rahmani2018}, and at $z\sim 1$ from \ovi\ and \mgii\ kinematics \citep[e.g.][]{Kacprzak2016, Bouche2016, Zabl2019}, \neww{thus subject to the inherent systematic limitations of using single beam probes}. In this \neww{context}, our result is the second positive detection of extended co-rotation in absorption in a galaxy at $z\approx 1$ after Paper~II.

With our tomographic data we \neww{find $\approx 85\%$} independent \mgii\ absorption data-points in a single galaxy system supporting this scenario. This is in contrast to the traditional studies based on quasar sightlines, that use typically few discrete data-points per galaxy in their ensembles. The gravitational arc-tomography has the advantage of reaching large distances with multiple, contiguous and independent background probes, which are essential for reducing the geometrical degeneracy of the model. Strong evidence of co-rotating extended \mgii\ (and \feii, \mgi) from multiple contiguous data-points has been previously reported in Paper~II, and the present system represents another such case. 

Interestingly, we are witnessing here the signatures of an extended disc, even though we are probing mostly positions aligned with the semi-minor axis of the targeted galaxy; but this is not surprising given the extent and inclination of \neww{G1}. This co-rotation on tens of kpc is consistent with the kinematics from the ISM of G1 over a few kpc (as probed by \oii\ emission), which implies a direct connection between the CGM and the ISM kinematics. 
%Theoretical work suggest that 
According to theory, 
this alignment of extended and inner discs may be a consequence of the gravitational and hydrodynamical interactions, e.g. from effective tidal torques that the galaxy applies to the inflowing material \citep[e.g.][]{Danovich2015,Stewart2017}.

Can pure rotation explain the large velocity dispersion generally observed? A possibility worth exploring is the effect of having an inclined thick disc of \mgii\ co-rotating gas that may produce an asymmetric absorption profile due to the combined effect of \new{the radial (along the disc) and perpendicular (to the disc) gas density profiles} \citep[e.g.][]{Prochaska1997, Ledoux1998}. We expect to have asymmetric ``edge-leading'' absorption where the \new{denser absorption component is accompanied by less dense absorption at higher(lower) absolute velocities for sightlines that are near(far) to the major-axis}, respectively \citep[e.g. see fig.~14 of][]{Prochaska1997}. We can test whether we observe a signature consistent with this in our dataset.

From the kinematics and morphology of the spiral arms seen in G1 \neww{(Figs.~\ref{fig:oii_2D}~and~\ref{fig:fov}, respectively)}, we can infer that the line-of-sight component of the angular momentum vector most likely points towards the observer \citep{Ho2020},\footnote{\new{Because spiral arms are generally expected to be trailing the disc rotation \citep[][but see \citealp{Vaisanen2008}]{Carlberg1985, Ho2020}.}\label{foot:vector}} implying the arc-N region is on the far-approaching part of the extended disc (and also far from the semi-major axis). Thus, from this model we would expect an asymmetry mostly towards the blue which is qualitatively consistent with what we observe from the higher resolution MagE data (\neww{Figs.~\ref{fig:mage_main}~and~\ref{fig:magefits};} see \S~\ref{sec:vdisp} and Appendix~\ref{sec:mage}). However, if this asymmetry is primarily due to the differential rotation effects of the putative thick disc then it would imply a large projected maximum velocity of $v\gtrsim 140$\,\kms, i.e. much larger than our inferred (de-projected) maximum rotational velocity of $v_{\rm rot}\approx 80$\,\kms. Of course, higher spectral resolution data should reveal even more complex absorption profiles than those seen \new{in our MUSE or MagE data}, including those signatures expected from a thick rotating disc at lower velocity differences, but here we conclude that the velocity dispersion {\it currently} observed should be mostly produced by other effects besides rotation \citep[e.g.,][Paper~II]{Charlton1998}.

\subsection{Turbulence and evidence of  \neww{galactic outflows}}\label{sec:inflow_outflow}

The high velocity dispersion observed for the \mgii\ absorption in all the regions indicates the presence of multiple non-resolved velocity  components. Large velocity deviations from a rotating disc model are expected in scenarios involving: turbulence, outflows, inflows, etc. In this section we explore whether we may be witnessing such kinematic signatures on top of that of \neww{rotation}.

\subsubsection{Evidence of turbulence}

Although most of the \mgii\ absorption centroids can be explained in the context of the ERD model, we can see from Fig.~\ref{fig:kinematics} that a few appear as ``kinematic outliers'' \neww{(\S~\ref{sec:vdisp}).} \neww{These} outliers are excellent candidates to be probing outflowing or inflowing gas clouds. There are $\noutliers$ such data-points, at $D\sim 15$--$20$\,kpc on both directions (arc-N and arc-S). In all the cases, their absolute velocities with respect to the systemic redshift are {\it larger} than what is predicted by the model, \neww{standing out} in Fig.~\ref{fig:kinematics} at $|\velm| \gtrsim 60$\,\kms. These spaxels present a mean $\bar{\vdispm}\approx 54 \pm 37$\,\kms (median $\sim 60$\,\kms), which is somewhat larger than the typical velocity dispersion seen in the rest of arc-N (see \S~\ref{sec:kinematics}). The large observed \vdisp\ are not driven by uncertainties in \vel, as these are typically $\lesssim 20$\,\kms. \neww{Thus,} we find typical values of $v_{\rm rot}/\vdispm \sim 1$ (consistent with the overall $v_{\rm rot}/\vdispm$ values) providing evidence of a
turbulent medium \citep[e.g.][]{Burkert2010}. 

The inferred maximum circular velocity for the dark matter halo of G1 is $v_{\rm max} = \vmaxhalo$\,\kms\ whereas the maximum rotation velocity for the gas is $v_{\rm rot, gas} = \vmaxoii$\,\kms. High redshift galaxies tend to show ratios of $v_{\rm max}/v_{\rm gas} = 1-5$ \citep[e.g.][]{Green2014}, thus G1 represents a typical example. Let us consider an exponential disc density profile \citep{Freeman1970} where the extra rotational support is given by pressure from turbulence where the dispersion is independent of radial distance or height \citep{Burkert2010}. In this context $v_{\rm rot, gas}^2 = v_{\rm max, h}^2 - 2\sigma^2 (r/r_d)$, where $r$ is the radial distance along the disc plane, and $r_d$ is the scale length of a disc exponential profile \citep{Burkert2010, Wang2010, Mortensen2021}. Thus, we would expect $r/r_d \sim 2$ for $\vdispm \approx 50-60$\,\kms, which implies a \mgii\ disc scale length of $r_d\sim 10-20$\,kpc considering $r\sim 20-40$\,kpc. This scale is several times larger than the effective radius of G1 measured from the stars and thus consistent with being in the CGM.  The presence of turbulence in the CGM on top of the ERD model would naturally explain most of the \mgii\ kinematic signatures we observe. 

The origin of turbulence is not yet fully understood and several astrophysical processes have been proposed to sustain it (which would otherwise decay quickly), including: star-formation feedback \citep[e.g.][]{Lehnert2009, Krumholz2010, Lehnert2013, Bacchini2020}, gravitational instabilities \citep[e.g.][]{Elmegreen2010, Burkert2010, Heigl2020}, ongoing cosmic accretion \citep[e.g.][]{Forster2006, Aumer2010}, and/or a combination of the above. The turbulence at high-redshift is most likely sustained by multiple mechanisms, although their relative importance is also expected to evolve with redshift \citep{Krumholz2018}.

At the redshift of G1, the Universe was $\sim 7$\,Gyr old, and only a few Gyrs after `cosmic noon'.\footnote{\neww{Defined at when the star-formation history of the Universe peaks, i.e. $z\sim 1-2$ \citep{Lilly1996, Madau1998, Madau2014}.}} Thus, one may expect star-formation feedback to be a relevant source of turbulence in the CGM (e.g. galactic wind outflows\new{; \citealt{Fielding2017}}). If this were the case, we should expect the turbulence signatures to be particularly enhanced towards the minor-axis of G1 \citep[e.g.][]{DeFelippis2020}. Considering that arc-N is probing the extended disc on the far-approaching side, a putative outflow would be preferentially seen as a blue-shifted component in the arc-N (and as a red-shifted component in arc-S). This scenario would then explain the absorption line asymmetry observed in our higher spectral resolution observations from MagE (see \S~\ref{sec:vdisp}), and can be tested from higher-$S/N$ observations of arc-S.

\subsubsection{Evidence of outflows}\label{sec:outflows}

Can the observed blue-shifted component seen in arc-N at $v_{\rm arcN, b} \approx -140$\,\kms\ originate from an outflow from G1? Taking into account the DtB absorption at $v_{\rm DtB, b} \approx -380$\,\kms\ that {\it must} be an outflow, one may consider if these two blue-shifted absorption features can be physically linked, e.g. modeled by a bi-conical galactic wind \citep[][]{Bouche2012, McQuinn2018,DeFelippis2020, Zabl2019, Zabl2020}. 

A key aspect supporting the presence of an outflow is the fact that the bluer components in arc-N are present in the full extent of the area probed by our MagE observations (see \S~\ref{sec:vdisp} and Appendix~\ref{sec:mage}). Let us assume a simple model where a galactic wind travels at constant speed $v_{\rm out}$ \citep[e.g.][]{Springel2003} with a half-cone angle of $\theta$ such that in projection is seen at $v_{\rm DtB, b}$ at $D\approx 0$\,kpc and at $v_{\rm arcN, b}$ at $D\approx 20$\,kpc on top of arc-N. We must then consider two cases: \new{a) $\theta> i$, and b) $\theta<i$, where $i$ is the disc inclination. In case a) we would expect $v_{\rm DtB, b} \approx v_{\rm out}$, while in case b) we would expect $v_{\rm DtB, b} \approx v_{\rm out}\cos(\theta - i)$.} In both cases the closest (and thus presumably denser) intersection of the bi-cone with the arc-N sightline is at the \new{same angle $\theta+i$ and thus $v_{\rm arcN, b} = v_{\rm out} \cos(\theta+i)$.} Given our values for $v_{\rm DtB, b}$, $v_{\rm arcN, b}$, and $i$, it is only possible to solve for case b), which gives $\theta\approx 25\degree$ and $v_{\rm out} \approx -400$\,\kms.

A half-cone angle of $\sim 25\degree$ is consistent with the higher incidence of \mgii\ absorption towards the minor axis of G1 (e.g. as seen in the second panel of Fig.~\ref{fig:mgii_2D}), and also consistent with previous theoretical \citep{Stewart2013, DeFelippis2020,Mitchell2020} and observational work \citep[e.g.][]{Bordoloi2011, Bouche2012, Kacprzak2012}.

For momentum driven winds, it has been estimated that velocities of galactic outflows are proportional to $2-3\times v_{\rm max, h}$ \citep[e.g.][]{Barai2013,Katsianis2017} which for G1 corresponds to $\approx 300-400$\,\kms. Thus, a simple bi-conical galactic outflow wind seems to be a valid scenario to explain the kinematic outliers to the ERD model.

Under this assumption then we should expect to see its signatures also as a redshifted component of order $\sim +140$\,\kms\ towards  arc-S. It is interesting to recall that all significant kinematic outliers in the arc-S are {\it redshifted} from the ERD model at absolute velocities \new{between} $\approx 60$--$120$\,\kms\  (Fig.~\ref{fig:kinematics}), thus consistent with the prediction. Unfortunately, our MagE \neww{data} of arc-S does not have enough $S/N$ to confirm or rule out this hypothesis.

Considering an outflow of $v_{\rm out} \sim 400$\,\kms, such galactic wind would be able to reach $0.5\rvirm$ in $\sim 0.2$\,Gyr which seems feasible considering the star-formation timescales. Although the current SFR of G1 is somewhat low ($\lesssim 1$\,\msun~yr$^{-1}$), our \ppxf\ analysis of the stellar light \neww{suggests} that the last important star-formation episode occurred typically $\gtrsim 1$\,Gyr into G1's past. In addition, and according to the inferred G1's sSFR (\ssfr), the galaxy may be just transitioning to become a quenched galaxy (i.e. sSFR $\lesssim 10^{-11}$\,yr$^{-1}$) unless new reservoir of cold gas is able to reach the ISM of G1 and ignite a new episode of star-formation (e.g. spiraling in from the observed ERD).

\subsubsection{\neww{Possibility of inflows}}

\neww{Although a galactic outflow may provide a viable explanation for the kinematic outliers seen towards arc-N, one may also consider the possibility of having inflows instead. The only qualitative requirement is to have a velocity vector component of the gas pointing towards the galaxy center, and there are certain geometries that allow this.} \neww{However, in those cases} one would need to explain the presence of galactic scale inflows seen mostly along the minor-axis of G1 with velocity offsets that are {\it larger} than the inferred maximum velocity of the extended disc, and comparable to that of the maximum circular velocity of the halo\neww{, which seems unlikely}. Thus, we disfavour this interpretation.

In summary, we conclude that \new{the most likely explanation for the absorption signatures observed \neww{around G1 towards \sgas}, is that} we are witnessing an extended and turbulent co-rotating disc \neww{where} the turbulence may be driven by a galactic outflow.

\subsection{Comparison with Mortensen et al. 2021}
\label{sec:mortensen}

% comparison with Mortensen+
\new{

\subsubsection{General comparison}

\citet{Mortensen2021} have recently presented a study of this same system using Keck/KCWI \citep{kcwi} observations of the arcs, at similar resolving power but covering a much smaller FoV. Our present work advances theirs by providing i) a determination of the systemic redshift of the absorbing galaxy, ii) partially-resolved emission and absorption-line kinematics, and iii) a systematic search for galaxies at G1's redshift on CGM scales.

Determining the systemic redshift of G1 
%based on the morpho-kinematical analysis to the [O~{\sc ii}] emission (see \S~\ref{sec:galpak}) 
together with the partially-resolved Mg~{\sc ii} absorption profile afforded by our Magellan/MagE data were critical for establishing the presence of both the extended co-rotating disk and the galactic-scale outflow in the CGM of G1 (see above). Similarly, the $9\times$ larger FoV allowed us to survey the galactic environment of G1 and establish its isolated nature.

On the other hand, our findings are consistent with theirs regarding the unresolved Mg~{\sc ii} absorption line measurements, as we both find that absorption towards arc-N has systematically lower redshift than absorption towards arc-S, with absolute differences being of order $\sim 100$\,\kms. Similarly, we also find consistency regarding the velocity dispersion of the absorbing gas, as we both report typical $\vdispm \approx 50$\,\kms. Considering these broad kinematical results in the context of the main properties of G1, both results point towards the conclusion that the \neww{CGM gas is} pressure supported (\S~\ref{sec:inflow_outflow}).

\subsubsection{On the robustness of impact parameters for different lens models}

Another important \neww{difference is} in regards of the \neww{lens models}. As can be seen from their fig.~3, the arc segments \neww{``move''} to the North-East \neww{after de-lensing}, whereas in our case \neww{they ``move''} mostly towards the South (see Fig.~\ref{fig:fov}). This difference is driven by a massive intervening galaxy cluster located $\sim 153\arcsec$ \neww{to} the South that is included in our model (see \S~\ref{sec:lens}; \citealt{Bayliss2011}), but \neww{not in theirs.}

However, as the de-lensed G1 and arc segments move {\it together} in one direction or another, the effect of this systematic difference has limited impact on the inferred de-lensed projected distances. We have quantified this effect as follows. We first used the tabulated values of spaxel positions (image plane) from \citet{Mortensen2021},\footnote{From their public pre-print version arXiv:2006.00006v1.}
to infer de-lensed impact parameters based on our lens model. We then directly compared these with their corresponding tabulated absorber-plane impact parameters. From this empirical comparison we find that at impact parameters $<20$\,kpc, the systematic differences are typically within $\sim10\%$, whereas at $>20$\,kpc our values are $\sim 20$\% \neww{larger. This} illustrates that lens models offer a relatively robust way to constrain impact parameters to within
$10\%$ for the most interesting scales under discussion ($\lesssim 20$\,kpc), despite the different assumption on the lens mass distributions.}

\subsection{Future prospects}

As a final remark, we note that the intrinsic properties of the galaxy studied here are in many respects similar to those of the galaxy studied in Paper~II, e.g. in terms of mass, SFR, redshift, inclination; yet the ratio of velocity dispersion and maximum velocity rotation $\vdispm/v_{\rm rot}$ appear qualitatively different between them: we see here a factor of $\sim 2\times$ larger $\vdispm/v_{\rm rot}$  than that of Paper~II for similar impact parameters. Although this apparent discrepancy may be explained by several factors, one possibility %that we consider worth exploring 
is that it may be a geometrical effect. Indeed, a key difference between both \neww{setups} is the alignment of the background gravitational arcs with respect to probed galaxies: while in Paper~II the arcs \neww{mostly probe} the CGM along the major axis, in this new case the arcs are more aligned with the minor axis. If similar galaxies at $z\sim 1$ have both extended discs {\it and} prominent galactic outflows, it seems reasonable to expect that signatures of the latter will be enhanced along their minor axis (as in the present case).
%it is reasonable to assume that signatures of the latter would be more readily
%apparent if we probe the CGM along the minor axis as in the present case. 
Thus, applying the gravitational-arc tomography to the study of azimuthal angle dependency of the CGM seems timely.

\section{Summary and conclusions}\label{sec:summary}

We have presented gravitational-arc tomography \neww{of} the circumgalactic medium (CGM) of an isolated galaxy at $z\approx 0.77$ towards the \sgas\ field. We based our results on a unique and powerful dataset composed by VLT/MUSE and Magellan/MagE spectroscopy, as well as \HST/WFC+ACS imaging. We could \neww{partially} resolve the kinematics of both the enriched CGM traced by \mgii\ absorption and the inner galaxy's interstellar medium (ISM) traced by \oii\ emission. The \hst\ and MUSE data allowed us to obtain the main physical parameters of the targeted galaxy (see Table~\ref{tab:G1}), as well as a precise lens model from which we properly reconstructed the geometry of the intervening galaxy and background arc-segments in the plane of the targeted galaxy. We have focused our present analysis on the kinematic signatures of the cool-warm enriched gas traced by \mgii\ absorption seen towards the arcs, and its connection to the galaxy.  Our experimental setup and main findings can be summarized as follows:

\begin{itemize}

    \item Forty-\new{two} independent spaxels seen towards three arc segments have enough $S/N$ to detect \mgii\ rest-frame equivalent widths of $\ewm^{\rm min} \gtrsim 0.3-0.4$\,\AA\ (Fig.~\ref{fig:fov}). The unique geometrical configuration of the arcs around the galaxy allowed us to perform a spatial mapping of the CGM over a wide range of azimuthal angles, covering the \new{major}-axis at impact parameters of $D \approx 10-30$\,kpc (on both sides), and the minor-axis at $D \approx 60$\,kpc (on one side). We detected \mgii\ absorption on \nhit\ independent spaxels all at  $D \lesssim 30$\,kpc (Figs.~\ref{fig:antonia}~and~\ref{fig:mgii_2D}),  from which we obtained the general kinematic trends of the enriched CGM.\smallskip
    
    \item Based on \oii\ emission from the galaxy we fit a morpho-kinematic rotating disc model, from which we \neww{infer} inclination, position angle, maximum rotation velocity, and kinematical centre of the galaxy's ISM (Fig.~\ref{fig:oii_2D}). We find \neww{them consistent with} those obtained by analysis of the stellar light based on \hst\ imaging. We extrapolated this kinematical model to the larger distances probed by the arcs (``extended rotating disc'', ERD), to compare it to \neww{the} \mgii\ absorption.  \smallskip

    \item \neww{We find that $\approx 85\%$} of the \mgii\ absorption centroids \neww{match} well the ERD model \neww{($\nERD$ out of $\nhit$ independent spaxels)}, providing evidence for the presence of an extended co-rotating \neww{disc} (Fig.~\ref{fig:kinematics}). This match implies a physical (e.g. dynamical) connection between the inner/central part of the galaxy (on distances $\lesssim 10$\,kpc) and the enriched gas material on distances of $\sim 20-40$\,kpc that is presumably in-falling \neww{into} an extended galaxy disc, as predicted by theoretical models of galaxy evolution.\smallskip

   \item The velocity dispersion \neww{of} \mgii\ absorption ($\vdispm \approx 30-100$\,\kms) is of the same order as the maximum gas rotational velocity inferred from the inner galaxy rotation ($v_{\rm rot, gas} \approx 80$\,\kms; Figs.~\ref{fig:mgii_2D}~and~\ref{fig:kinematics}), providing evidence for the existence
    of a turbulent and pressure-supported CGM. As we are probing the CGM mostly along the minor-axis to the galaxy, a galactic outflow scenario is a reasonable explanation for the origin of the large velocity dispersion we observe \new{in the absorption profiles}. An outflow with $|v_{\rm out}| = 400$\,\kms\ and half-cone angle of $\theta\approx 25$\degree\ can explain not only the few kinematic outliers to the ERD \new{model}, but also the asymmetries in the \mgii\ absorption line profiles observed in our \new{Magellan/MagE higher spectral resolution data against both the arc-N and the galaxy} (Fig.~\ref{fig:mage_main}).\smallskip
    
\end{itemize}

\neww{These} findings highlight the power of our technique to unveil the signatures of the CGM at intermediate redshifts. In the present case, we are undoubtedly witnessing key aspects of the baryon cycle of metals in the CGM of a single isolated galaxy, namely extended co-rotation and a galactic outflow. Gravitational arc-tomography \neww{thus} provides a unique new view of the complex and dynamic CGM of galaxies around cosmic noon.

\section*{ACKNOWLEDGEMENTS}

\new{We thank the anonymous referee for their constructive criticism and comments that \neww{substantially} improved the quality of this manuscript. We thank K.~Rubin and S.H.~Ho for useful comments and discussions. N.T., S.L, A.F., N.R. and F.C. acknowledge support by FONDECYT grant 1191232. E.J.J. acknowledges support from FONDECYT Postdoctoral Fellowship Project No.\,3180557 and the BASAL Center for Astrophysics and Associated Technologies (PFB-06). G.D. acknowledges support from CONICYT project Basal AFB-170002.} This paper is based on observations collected at the European Southern Observatory under ESO programme(s)
0101.A-0364(A) (PI Lopez). This paper \neww{also} includes data
gathered with the $6.5$ meter Magellan Telescopes located at Las Campanas Observatory, Chile; the Magellan/MagE observations were carried out as part of program CN2019A-36 (PI Tejos).

This paper is partly based on observations made with the NASA/ESA {\it Hubble Space Telescope} obtained from the Space Telescope Science Institute, which is operated by the Association of Universities for Research in Astronomy, Inc., under NASA contract NAS 5-26555. These observations are associated with programs 
GO~15378 (PI~Bayliss), 
GO~12116 (PI~Ebeling), and
GO~12368 (PI~Morris).

We thank contributors to SciPy\footnote{\url{http://www.scipy.org}},
Matplotlib\footnote{\url{http://www.matplotlib.sourceforge.net}},
Astropy\footnote{\url{http://www.astropy.org}
  \citep{astropy}}, Linetools\footnote{\url{https://linetools.readthedocs.io/en/latest/} \citep{linetools}}, \new{ESO/MIDAS}\footnote{\url{https://www.eso.org/sci/software/esomidas/}}, the Python programming
language\footnote{\url{http://www.python.org}}, and the free and
open-source community. This research has also made use of the NASA's
Astrophysics Data System
(ADS)\footnote{\url{http://ned.ipac.caltech.edu}}.

\section*{DATA AVAILABILITY}
The data from the Hubble Space Telescope used in this paper may be accessed
from the Mikulski Archive for Space Telescopes (MAST)\footnote{\url{https://archive.stsci.edu/}}, while that from the European
Southern Observatory (ESO) may be accessed from the ESO Archive\footnote{\url{http://archive.eso.org/}}, using the relevant program IDs as given in the Acknowledgements section. The Magellan/MagE data will be shared on reasonable request to the corresponding author.

\bibliographystyle{mnras}
\bibliography{g1-j1226}

%appendix
\input{appendix}
\bsp
\label{lastpage}

\end{document}

%% file: abstract_v2.tex
%[abstract TBD] 

We present gravitational-arc tomography of the cool-warm enriched circumgalactic medium (CGM) of an isolated galaxy (``G1'') at $z \approx 0.77$. Combining VLT/MUSE adaptive-optics and Magellan/MagE echelle spectroscopy we obtain partially-resolved kinematics of Mg~{\sc ii} in absorption and \oii\ in emission. The unique arc configuration allows us to probe 42 spatially independent arc positions transverse to G1, plus 4 positions in front of it. The transverse positions cover G1's minor and major axes at impact parameters of $\approx 10-30$\,kpc and $\approx 60$\,kpc, respectively. We observe a direct kinematic connection between  the cool-warm enriched CGM (traced by Mg~{\sc ii}) and the interstellar medium (traced by [O~{\sc ii}]). This provides strong evidence for the existence of an extended disc that co-rotates with the galaxy out to tens of kiloparsecs. The Mg~{\sc ii} velocity dispersion ($\vdispm \approx 30-100$\,\kms, depending on position) is of the same order as the modeled galaxy rotational velocity ($v_{\rm rot} \approx 80$\,\kms), providing evidence for the presence of a turbulent and pressure-supported CGM component. We regard the absorption to be modulated by a galactic-scale outflow, as it offers a natural scenario for the observed line-of-sight dispersion and asymmetric profiles observed against both the arcs and the galaxy. An extended enriched co-rotating disc together with the signatures of a galactic outflow, are telltale signs of metal recycling in the $z\sim 1$ CGM.

%% file: fig_muse.tex
\begin{figure}
%\begin{minipage}{1.7\columnwidth}
\begin{minipage}{1.\columnwidth}
\includegraphics[width=\columnwidth]{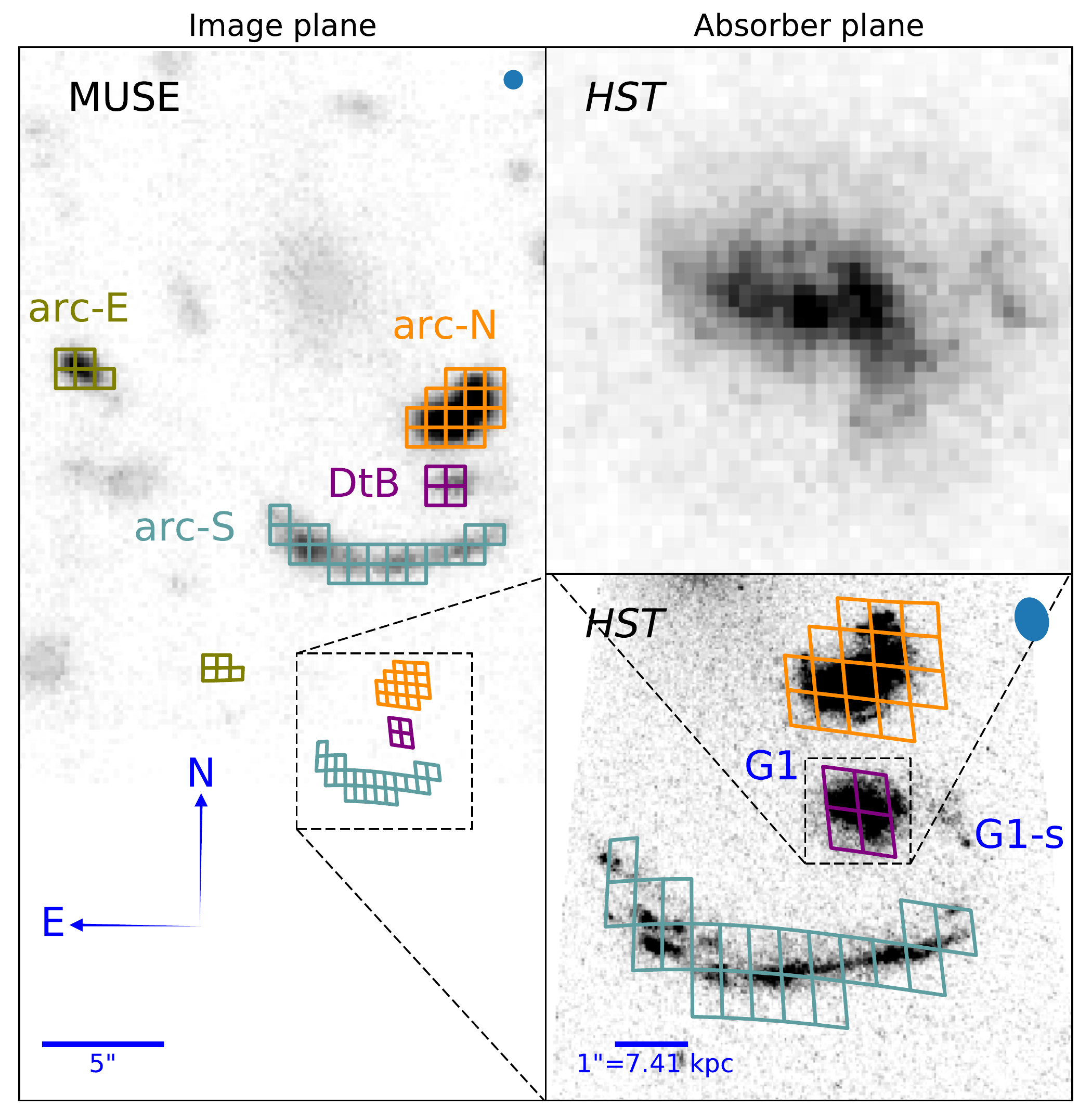} 
\end{minipage} %{1\columnwidth}
\caption{
{\it Left:} \neww{Subsection of the MUSE field-of-view towards \sgas\ centred at the arc continuum near the Mg~{\sc ii} absorption. The three arc segments and the ``down-the-barrel'' binned spaxels are indicated (\S~\ref{sec:arcs}) and have been color-coded for clarity here and in 
Figs.~\ref{fig:antonia}~and~\ref{fig:kinematics}.  Native MUSE spaxels are binned $4\times 4$ ($0.8\arcsec\times0.8\arcsec$) to obtain uncorrelated signals, considering that the ground-based observations have an effective PSF FWHM of $0.7\arcsec$ (indicated by the blue circle). Also shown is the re-mapping of the grid into the absorber plane (\S~{\ref{sec:lens}}). {\it Right bottom:} Zoom-in around G1 in the reconstructed absorber plane showing the de-projected {\it HST} image and grid. The background image
corresponds to the de-lensed {\it HST}/F814W. ``G1'' indicates the absorbing galaxy, and ``G1-s'' indicates a star-forming region and/or a small satellite to the West of G1 (\S~\ref{sec:emission}~and~Fig.~\ref{fig:oii_2D}). The effective PSF at the position of G1 is indicated by the blue ellipse.
{\it Right top:} Further zoom-in into G1 in the absorber plane with a different level-cut to emphasize the spiral arms. }
%(the latter was used to
%fit the emission-line cube). 
%{\bf Perhaps indicate spaxel numbering, and refer to the Fig.
%with spectra}.
\label{fig:fov}
}
\end{figure}

%% file: galaxy_g1.tex
\section{G1 emission properties}\label{sec:emission}

%table with G1 properties
\input{table_G1}

\neww{We use the \HST\ and MUSE data to} infer the main properties of G1, which are summarized in Table~\ref{tab:G1}.

\subsection{Mass and luminosity}

We used the \bagpipes\ software \citep[][]{Carnall2018} to estimate G1's stellar mass from both the \HST\ photometry and the MUSE continuum spectra. The \HST\ magnitudes were measured in the image plane (corrected by Galactic extinction; see Table~\ref{table:g1_hst} \new{in the Appendix}), and a G1's $1$-d spectrum \neww{(excluding G1-s) was optimally} extracted using the \pymuse\ software, also in the image plane. \new{The stellar population models used by \bagpipes\ were the 2016 versions of those from \citet{BC03}, assuming the stellar initial mass function (IMF) of \citet{Kroupa2002}. Star-formation histories were modeled as double-power-laws \citep[e.g.][]{Behroozi2013}.}
We obtained a stellar mass of $M_*=10^{\massstars}$\,\msun\ after correcting by the  magnification factor $\mu=2.7$. This stellar mass implies a dark matter halo mass of $M_{\rm h}=10^{\halomassstars}$\,\msun\ based on the abundance matching relation of \citet{Moster2010}.

We estimated the galaxy virial radius as $R_{\rm vir} = [3M_{\rm h}/(4 \pi 200 \rho_{\rm c}(z))]^{1/3}$ and the maximum circular velocity as $v_{\rm max} = [10 M_{\rm h} G H(z) ]^{1/3}$ \citep[e.g.][]{Mo1998}, where $G$ is the gravitational constant, and $\rho_{\rm c}(z)$ and $H(z)$ are the critical density of the Universe and Hubble parameter at redshift $z$, respectively. For G1 we obtained $R_{\rm vir}= \rvir$\,kpc and $v_{\rm max} = \vmaxhalo$\,\kms. 

\new{We estimated a $B$-band luminosity of
$L_B=(9.1 \pm 0.7)\times 10^{9}L_{\odot}$
based on the \HST/814W photometry after correcting for the magnification, corresponding to $L_B=\LumB$ at $z=0.77$ \citep{Willmer2006}.}

\input{fig_oii}

% Morphology
\subsection{Morphology}

We used the \galfitm\ software \citep[v1.2.1][]{Haeussler_2013, Vika_2013} \neww{to} estimate the morphology of G1 from the stellar light. \new{In this case, we de-lensed the broad band \HST\ images to the absorber plane before running it through \galfitm. \galfitm\ convolves the model image with a PSF profile provided by the user, and so these PSF profiles for each band were de-lensed in the same way. Modelling the galaxy with a single S\'ersic profile resulted in excess light in the residual image at the core of the galaxy, indicating a poor fit in this \neww{central} region. The fits were then repeated using a central PSF component in addition to the S\'ersic profile, which resulted in a better fit to the delensed image.} We set the Chebychev polynomials to be of order $1$ (i.e. constant with wavelength) as it provided consistent structural parameters among the different bands, specifically we obtained: the effective radius ($R_e=\eradstars$\,kpc), the S\'ersic index ($n=\sersic$), the position angle (PA$_*= \pastars$\degree), and the inclination ($i_{*}=\inclstars$\degree).

After subtracting the smoothed light of the galaxy we see clear residual features in the stellar light, consistent with spiral arms and/or tidal tails following a close approach or a merger with another galaxy. These features are readily evident in the \neww{right-panels} of Fig.~\ref{fig:fov}. \neww{We also note on the presence of prominent [O~{\sc ii}] emission from G1-s (see Fig.~\ref{fig:oii_2D} and text below); altogether, these features may represent} the signatures of a less-massive galaxy being accreted by G1 that may have undergone the first or second approach.

\subsection{Star-formation rate}
% star-formation-rate
We estimate G1's star-formation-rate (SFR) primarily from the extracted MUSE $1$-d spectrum. We first modeled the stellar continuum contribution using the \ppxf\ software \citep[][]{Cappellari2017,Cappellari2004} masking out emission lines. This stellar continuum was then subtracted to the observed spectrum from which we measured emission line fluxes \new{(see Figure~\ref{fig:ppxf} in the Appendix)}.  Here we focus only in the [O~{\sc ii}] \neww{doublet}, primarily motivated by comparison with other similar studies (e.g. Paper~I and Paper~II). We obtained \neww{a total [O~{\sc ii}]\ emission flux of $f_{\rm [OII]} =$\fluxoii\ }(corrected by dust extinction and magnification), implying a SFR$_{\rm [OII]}$=\sfroii\ \new{(\citealt{Kennicut1998}; assuming a \citealt{Kroupa2002} stellar IMF)}. From the \bagpipes\ analysis we also estimate a SFR from the continuum spectra (hence probing larger timescales) of SFR$_{\rm SED}$=\sfrsed. Both estimates are within the same order of magnitude, and their errors are dominated by systematic uncertainties.  \new{Our derived SFR values are lower than that of  \citet[][SFR$=10 \pm 5$\,\msun\,yr$^{-1}$ obtained from SED fitting]{Mortensen2021}. One possibility to explain this discrepancy is that G1 is partially obscured by dust, making our estimations lower limits; however, given the large systematic errors involved in both derivations, the true value may lie between both estimates.}
Finally, given the stellar mass inferred above, we estimate a characteristic specific star-formation-rate of \neww{sSFR$=$\ssfr,} which makes G1 \new{broadly consistent with the main-sequence at $z\approx 0.8$ \citep[e.g.][]{Ilbert2013,Whitaker2014, Speagle2014,Furlong2015}. Using a sSFR=$5 \times 10^{-11}$\,yr$^{-1}$ cut to separate star-forming versus quenched galaxies at $z=0.8$, G1 may well be in the transition phase to become quenched.}

% \subsection{Morpho-kinematics from [O~{\sc ii}] emission} \label{sec:galpak}
\subsection{Morpho-kinematics}\label{sec:galpak}

\neww{G1's [O~{\sc ii}] emission} appears partially resolved in the MUSE datacube, which allow us to perform a $3$-dimensional morpho-kinematic analysis (see Fig.~\ref{fig:oii_2D}). We used the \galpak\ software \citep[v1.27;][]{Bouche2015} to fit \neww{a rotating} disc with the following parameters: an exponential flux profile, an arctan rotation \new{curve}, and a Gaussian thickness profile. In a similar fashion as in Paper~II, \neww{the input is} a de-lensed \oii\ emission (continuum subtracted) datacube of G1 \neww{(excluding G1-s)} and \neww{an} effective de-lensed PSF. The delensed datacube was resampled on a $0.1\times0.1$\,arcsec$^2$ grid, which matches better the expected size of the delensed MUSE native spaxels in the absorber plane. We run \galpak\ until convergence and found a satisfactory kinematical solution. The main  parameters of this model are: the maximum rotation velocity ($v_{\rm max, gas} =\vmaxoii$\,\kms), the velocity dispersion ($\sigma_{\rm v, gas} = \vdispoii$\,\kms), 
the inclination ($i_{\rm gas} = \incloii$\degree), the position angle (PA$_{\rm gas} = \paoii$\degree), and the turn-over radius ($r_t = \troii$\,kpc). We also obtained a kinematical center of the model which we used to define the systemic redshift of G1 and its center in the absorber plane used in the subsequent analysis: \new{$z_{\rm sys}=\zgal \pm 0.00002$} and (R.A., Decl.) = (\ragalpak,\decgalpak) J2000 (different than the image plane coordinate given in Table~\ref{tab:G1}). 

The bottom-right panel of Fig.~\ref{fig:oii_2D} shows the velocity field of this model for individual de-lensed unbinned native MUSE \neww{spaxels,} which can be contrasted with the actual observations of \oii\ velocity centroids \new{(from individual double Gaussian fits)}
in the top-right panel. We emphasize here that the latter come directly from the MUSE unbinned data, whereas the former was obtained from resampling \neww{quadrilateral} areas (i.e. those of the native MUSE spaxels delensed to the absorber plane) into perfect squares as input of the \galpak\ modeling. The observed good qualitative agreement between the morpho-kinematical model and the observed centroids \neww{for G1} indicate that the solution is sensible, despite the oversampling. \neww{We note that the model overestimates the recession velocities at G1-s, indicating that this region is not sharing the exact same kinematics as G1.}

The morphological parameters of \neww{the disc model} are consistent with those obtained from the stellar light (see above). In contrast to PA$_*$ that is solely determined from the apparent ellipticity of the galaxy stellar light, PA$_{\rm gas}$ is robust despite the \neww{low} inclination because it is mostly determined by the direction of the largest gradient in the velocity field which is well constrained by the observations.

%% file: table_G1.tex
\begin{table}
\caption{G1 properties}\label{tab:G1}
\centering
\resizebox{1.05\columnwidth}{!}{
\begin{tabular}{lr}
\hline
\hline
% \multicolumn{2}{c}{\it From broad-band imaging and \oii\ emission (see \S~\ref{sec:emission})}\\
% \multicolumn{2}{c}{\it Location}\\
%\hline
Redshift & $z_{\rm sys}=\zabs \pm 0.00002$ \\
Right Ascension (J2000)$^a$ & R.A. = \ra \\
Declination (J2000)$^a$     & Dec. = \dec \\
% \multicolumn{2}{c}{\it Morphological parameters}\\
Inclination angle (stars)$^b$ & $i_*=\inclstars\,\degree$  \\
Inclination angle (gas)$^b$ & $i_{\rm gas}=\incloii\,\degree$\\
Position angle (stars)$^b$ & PA$_*=\pastars\,\degree$  \\
Position angle (gas)$^b$ & PA$_{\rm gas}=\paoii\,\degree$\\
Effective radius (stars)$^b$ & $R_e=\eradstars$\,kpc\\
% Half-light radius (gas) & $r_{1/2}= \rgas$\,kpc \\ unconstrained
Sersic index (stars)$^b$ & $n=\sersic$\\
%$B$-band absolute magnitude$^c$ &  $M_B=\MagB$\\
\new{$B$-band luminosity$^c$}&$L_B=\LumB$ \\
Stellar mass$^c$ & $\log(M_*/\msunm)$=$\massstars$\\
Halo mass (from M$_*$) &  $\log(M_h/\msunm)$=$\halomassstars$\\
Virial radius (from M$_{\rm h}$) & $R_{\rm vir}=\rvir$\,kpc\\
Maximum circular velocity (from M$_{\rm h}$) & $v_{\rm max} = \vmaxhalo$\,\kms\\
%SF-efficiency$^{b, c, d}$&SFE=$3.5\pm 1.2\times10^{-10}$\,yr$^{-1}$\\
Maximum rotation velocity (gas)$^{b,e}$ & $v_{\rm rot, gas} = \vmaxoii$\,\kms\\
%Velocity dispersion (stars) & $\sigma_{\rm v, *} = \vdispstars$\,\kms\\
Velocity dispersion (gas)$^{b,e}$ & $\sigma_{\rm v, gas} = \vdispoii$\,\kms\\
Turnover radius (gas)$^{b,e}$ & $r_{t}=\troii$\,kpc  \\
\oii~flux$^{c,d} $ & $f_{\rm [OII]}=$ \fluxoii \\
Star-formation rate from \oii$^{c,d}$ &SFR$_{\rm [OII]}$ = \sfroii \\
Star-formation rate from SED$^{c,d}$ & SFR$_{\rm SED}$=\sfrsed \\
Specific SFR$^{c, d}$ & sSFR = \ssfr\\
% Older stellar age ($\approx 80\%$ of $M_{*}$)  & $\tau_{\rm old} = \agei$\,Gyr \\
% Younger stellar age ($\approx 20\%$ of $M_{*}$)  & $\tau_{\rm young} = \ageii$\,Gyr \\
%Dynamical mass at $r_{1/2}$ & M$_{\rm dyn}\approx 2\times10^{10}$\,M$_{\astrosun}$\\
% Halo mass (from dynamics) & $\log(M^{\rm dyn}_h/\msunm) = 10.83 \pm 0.04$ \\
\hline
\end{tabular}
}
\\
\vspace{1ex}
\begin{minipage}{0.9\columnwidth}
\footnotesize{
\raggedright {\bf Notes:}\\
$^a$ In the image plane.\\
$^b$ In the absorber plane (delensed). \\
$^c$ De-magnified quantity using $\mu=2.7$ (see \S~\ref{sec:lens}).\\
$^d$ Corrected by dust extinction.\\
$^e$ Defined from the arctan rotation curve:\\ $v(r)=v_{\rm rot}\frac{2}{\pi} \arctan(r/r_t)$.
}
\vspace{2ex}
\end{minipage}
\end{table}

%% file: fig_oii.tex
\begin{figure}
\begin{minipage}{1\columnwidth}
%\flushright
   \includegraphics[width=1\textwidth]{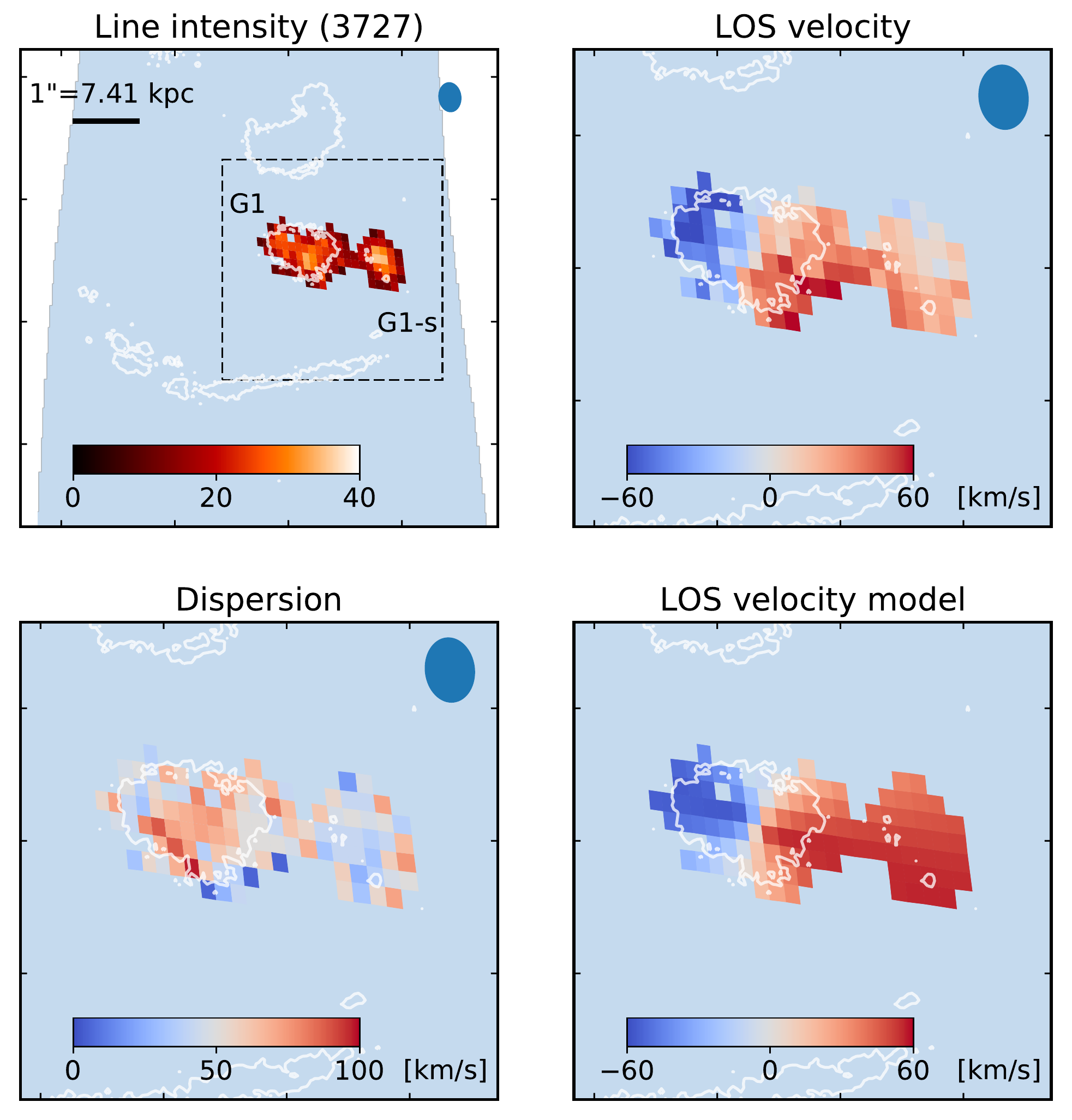}
\end{minipage}
\caption{\oii\ emission maps of G1 and G1-s in the delensed absorber plane \new{using native MUSE spaxels}. From top left to bottom right: \neww{observed} $\lambda 3727$ line intensity in units of $10^{-20}{\rm erg~s}^{-1}{\rm cm}^{-2}$~\new{per pixel}; \neww{observed line-of-sight velocity} with respect to \zsys=\zabs; velocity dispersion; \neww{G1's} modeled \neww{line-of-sight} velocity (see \S~\ref{sec:galpak}). \new{The latter 3 panels zoom in into G1, as indicated by the inset in the upper left panel.}  \neww{White contours correspond to the \hst/F814W image}. The de-lensed effective PSF is represented by the blue ellipse on the top-right.
%of each panel (excepting that of the model) as reference. 
\neww{We note that the G1 model overestimates the recession velocities of G1-s.}}
\label{fig:oii_2D}
\end{figure}

%% file: fig_mgii_spectra.tex
\begin{figure*}
\begin{minipage}{2\columnwidth}
\centering
\includegraphics[width=1.00\textwidth,trim={0cm 0cm 0.0cm 0cm}]{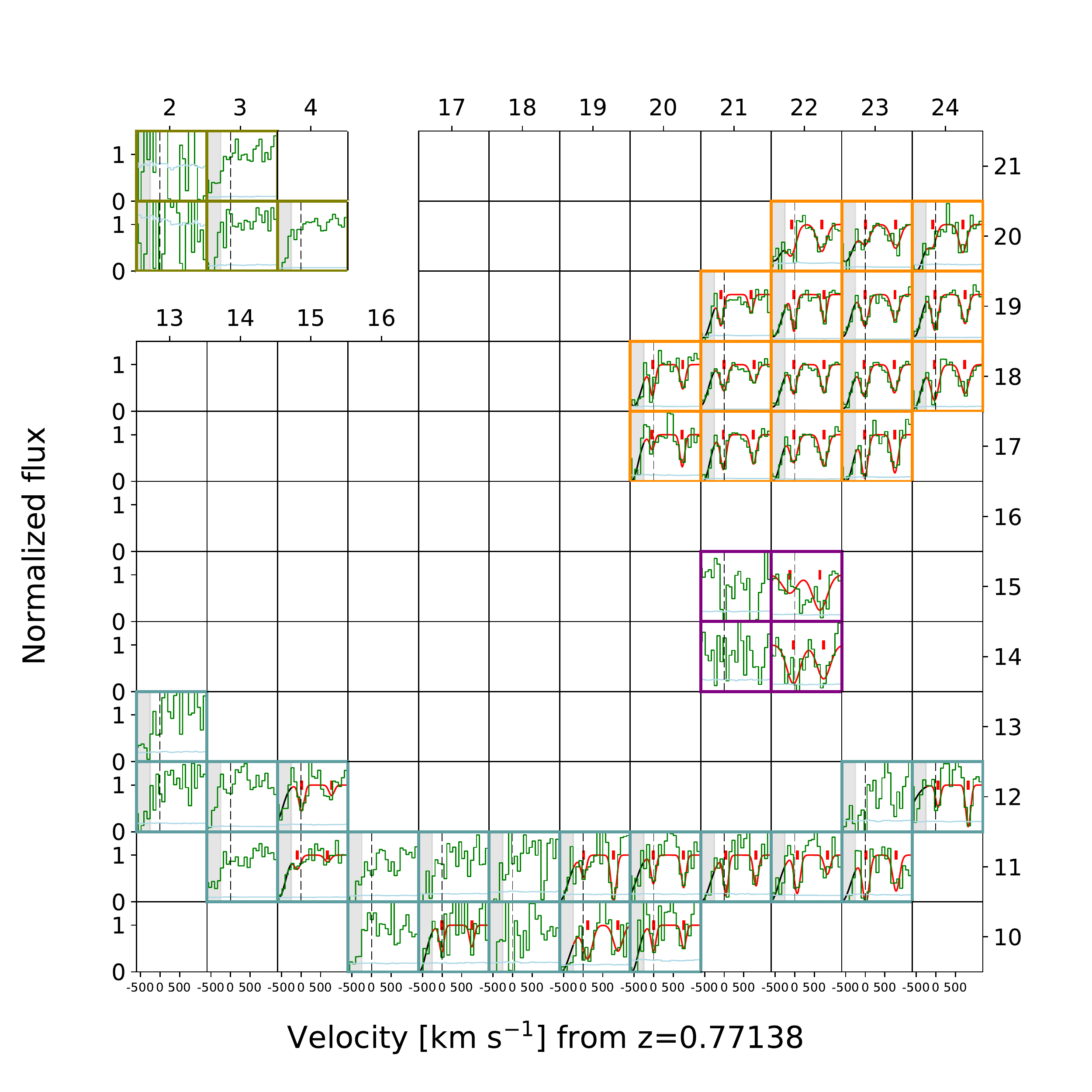}
\end{minipage}
% \caption{{\it Left panel}:
\caption{\neww{Map of} \mgii\ absorption \neww{spectra in} each MUSE binned spaxel (\S~\ref{sec:binning}) \neww{at $R\sim 2\,000$} ordered by their relative positions in the image plane \neww{and coloured as in Fig.~\ref{fig:fov}}. Each square shows the normalized \neww{flux and its uncertainty (green and light-blue histograms, respectively),} and their \mgii\ \new{doublet} fits (solid red lines). \neww{The} individual $x$-axes correspond to rest-frame velocities of the \mgii$\lambda 2796$ line with respect to $z_{\rm sys}=\zabs$ (marked by vertical black dashed lines).
%and the \neww{$y$-axes are} the normalized flux. 
\new{The centroids of the \neww{fits} are marked by vertical red lines.} The numbers around the top and right sides of the map define the spaxel coordinates $(x,y)$ (e.g. used in Table~\ref{tab:mgii}). The partially blended \Siii\ line at the redshift of the arcs is marked by the grey regions and \neww{black} lines.
%{\it Right panel}: A combined spectrum of G1 for \mgii\ absorption (top) and \oii\ emission (bottom), in a region defined as XXX (see \S~\ref{sec:vdisp}). All rest-frame velocities are given relative to $z_{\rm abs}=\zabs$ (vertical black dashed lines).
%\CL{We need to document the presence of the emission line in between the MgII absorption lines in the arc spectra as well as the presence of SiII absorption bluewards of it, in both the figure and main text.}
}
\label{fig:antonia}
\end{figure*}

%% file: fig_mgii_2D.tex
\begin{figure*}
\begin{minipage}{2.0\columnwidth}
%\flushright
   \includegraphics[width=1.\textwidth]{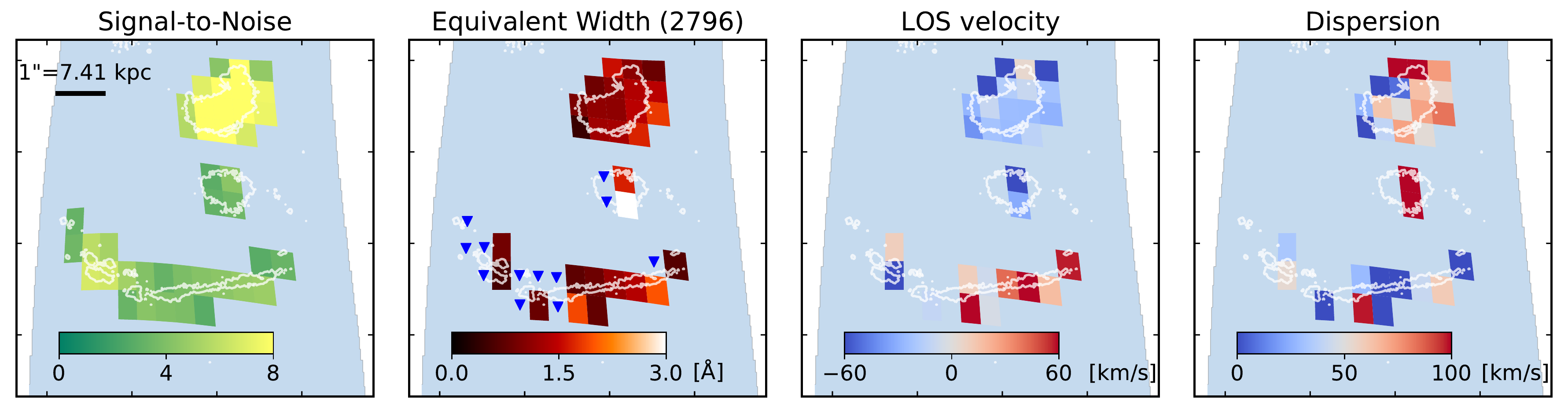}
\end{minipage}
\caption{\neww{Observed} \mgii\ absorption maps in the absorber plane with binned MUSE spaxels (see \S~\ref{sec:binning}). From left to right: continuum S/N at \mgii; rest-frame equivalent width (\ew); \neww{line-of-sight velocity} with respect to $z_{\rm sys}=\zabs$ (\vel); and \neww{line-of-sight} velocity dispersion ($\sigma$). The blue triangles in the second panel \neww{indicate} non-detections. \neww{White contours correspond to the \hst/F814W image.}
%We emphazise that the absorber plane our spaxels are rhomboids.
}
\label{fig:mgii_2D}
\end{figure*}

%% file: table_mgii_nt.tex
\begin{table*}
\begin{minipage}{0.6\textwidth}
\caption{\mgii\ absorption measurements in the MUSE spaxels}
\label{tab:mgii}
\end{minipage}
\centering
\begin{tabular}{c c c c c c c}
\hline\hline
Region  & Spaxel   & $D$   & W$_{0}$2796 & $v$           & $\sigma$      & ERD? \\
        & (x,y)    & (kpc) & (\AA)       & (km s$^{-1}$) & (km s$^{-1}$) & (y/n) \\
 (1)    &  (2)     & (3)   &  (4)        & (5)           & (6)           &  (7)   \\ 
\hline
arc-N&(20,17)&$-12.7$&$0.45 \pm 0.2$&$-39.7 \pm 12.0$&$0.0 \pm 13.9$&y\\ 
arc-N&(21,17)&$-11.1$&$1.26 \pm 0.13$&$-22.0 \pm 5.1$&$40.0 \pm 5.4$&y\\ 
arc-N&(22,17)&$-10.0$&$1.31 \pm 0.12$&$-25.4 \pm 5.4$&$73.0 \pm 5.6$&y\\ 
arc-N&(23,17)&$-9.8$&$1.7 \pm 0.2$&$-13.3 \pm 6.4$&$52.3 \pm 6.7$&y\\ 
arc-N&(20,18)&$-15.9$&$0.97 \pm 0.23$&$-27.6 \pm 9.9$&$26.7 \pm 10.8$&y\\ 
arc-N&(21,18)&$-14.5$&$1.13 \pm 0.12$&$-9.8 \pm 6.1$&$62.3 \pm 6.4$&y\\ 
arc-N&(22,18)&$-13.6$&$1.12 \pm 0.06$&$-24.8 \pm 2.9$&$50.7 \pm 3.1$&y\\ 
arc-N&(23,18)&$-13.4$&$1.46 \pm 0.09$&$-25.2 \pm 4.0$&$72.3 \pm 4.4$&y\\ 
arc-N&(24,18)&$14.0$&$1.83 \pm 0.28$&$-28.8 \pm 10.6$&$83.4 \pm 11.5$&y\\ 
arc-N&(21,19)&$-17.8$&$0.87 \pm 0.25$&$-82.5 \pm 12.5$&$0.0 \pm 12.3$&n\\ 
arc-N&(22,19)&$-17.0$&$1.09 \pm 0.11$&$-16.5 \pm 4.3$&$7.9 \pm 4.6$&y\\ 
arc-N&(23,19)&$-16.8$&$1.39 \pm 0.11$&$-9.3 \pm 4.7$&$64.4 \pm 4.9$&y\\ 
arc-N&(24,19)&$-17.1$&$1.46 \pm 0.19$&$-21.6 \pm 7.3$&$55.0 \pm 7.4$&y\\ 
arc-N&(22,20)&$-20.3$&$1.58 \pm 0.62$&$-78.3 \pm 24.7$&$103.5 \pm 29.8$&y\\ 
arc-N&(23,20)&$-20.0$&$1.09 \pm 0.26$&$4.7 \pm 14.7$&$100.8 \pm 16.7$&y\\ 
arc-N&(24,20)&$-20.4$&$0.84 \pm 0.36$&$-79.2 \pm 18.8$&$74.3 \pm 20.5$&n\\ 
arc-S&(16,10)&$-24.7$&$<0.5$&--&--&--\\ 
arc-S&(17,10)&$-22.8$&$0.83 \pm 0.38$&$-11.1 \pm 19.2$&$0.0 \pm 19.2$&y\\ 
arc-S&(18,10)&$-21.3$&$<0.57$&--&--&--\\ 
arc-S&(19,10)&$20.2$&$1.92 \pm 0.64$&$117.7 \pm 18.0$&$98.4 \pm 23.1$&n\\ 
arc-S&(20,10)&$19.6$&$0.78 \pm 0.45$&$-3.6 \pm 24.9$&$0.0 \pm 23.1$&y\\ 
arc-S&(14,11)&$-26.3$&$<0.27$&--&--&--\\ 
arc-S&(15,11)&$-23.8$&$0.54 \pm 0.25$&$-92.4 \pm 27.0$&$53.4 \pm 28.5$&y\\ 
arc-S&(16,11)&$-21.5$&$<0.38$&--&--&--\\ 
arc-S&(17,11)&$-19.3$&$<0.46$&--&--&--\\ 
arc-S&(18,11)&$-17.5$&$<0.61$&--&--&--\\ 
arc-S&(19,11)&$-16.0$&$0.73 \pm 0.28$&$10.2 \pm 10.8$&$28.5 \pm 11.0$&y\\ 
arc-S&(20,11)&$15.1$&$0.86 \pm 0.32$&$-7.0 \pm 13.1$&$0.0 \pm 12.8$&y\\ 
arc-S&(21,11)&$14.9$&$1.2 \pm 0.38$&$42.3 \pm 11.3$&$0.0 \pm 13.6$&y\\ 
arc-S&(22,11)&$15.4$&$1.41 \pm 0.39$&$63.3 \pm 12.3$&$42.3 \pm 14.6$&n\\ 
arc-S&(23,11)&$16.7$&$1.99 \pm 0.41$&$17.9 \pm 10.6$&$60.2 \pm 11.8$&y\\ 
arc-S&(13,12)&$-27.0$&$<0.49$&--&--&--\\ 
arc-S&(14,12)&$-24.2$&$<0.3$&--&--&--\\ 
arc-S&(15,12)&$-21.5$&$0.9 \pm 0.41$&$10.2 \pm 21.1$&$33.6 \pm 22.8$&y\\ 
arc-S&(23,12)&$12.3$&$<0.64$&--&--&--\\ 
arc-S&(24,12)&$14.5$&$0.67 \pm 0.4$&$57.3 \pm 15.5$&$0.0 \pm 15.6$&y\\ 
arc-S&(13,13)&$-25.7$&$<0.54$&--&--&--\\ 
arc-E&(2,20)&$-62.0$&$<0.37$&--&--&--\\ 
arc-E&(3,20)&$-58.1$&$<0.2$&--&--&--\\ 
arc-E&(4,20)&$-54.4$&$<0.42$&--&--&--\\ 
arc-E&(2,21)&$-63.2$&$<0.25$&--&--&--\\ 
arc-E&(3,21)&$-59.4$&$<0.2$&--&--&--\\ 
DtB&(21,14)&$-3.1$&$<0.68$&--&--&--\\ 
DtB&(22,14)&$2.0$&$3.29 \pm 1.05$&$-31.6 \pm 40.7$&$157.7 \pm 35.9$&--\\ 
DtB&(21,15)&$-4.1$&$<0.56$&--&--&--\\ 
DtB&(22,15)&$-2.1$&$1.68 \pm 0.38$&$-126.9 \pm 21.9$&$170.8 \pm 19.0$&--\\ 

\hline
\end{tabular}\\
\vspace{1ex}
\begin{minipage}{0.6\textwidth}
\raggedright {\bf Notes:} 
(1) Region of the spaxel (see \S~\ref{sec:arcs}); 
(2) Spaxel number to those in Fig.~\ref{fig:antonia}; 
(3) Impact parameter of the spaxel to G1 in the
absorber plane; positive values are arbitrarily defined 
to be on the receding side of the velocity model, and negative 
values otherwise (see \S~\ref{sec:ERDcomp}); 
%(4) Azimuthal angle of the spaxels with respect to the semi-major axis of G1 at PA$=67$\degree' 
(4) Rest-frame equivalent width of the \mgii\,$\lambda$2796 line (non-detections are given as $2\sigma$ upper limits); 
(5) Rest-frame velocity centroid of the \mgii\,$\lambda$2796 line with respect to the systemic redshift, $z_{\rm sys}=\zabs$;
(6) Rest-frame velocity dispersion of the \mgii\,$\lambda$2796 line (deconvolved by the LSF); 
(7) Whether the velocity centroid lies within $3\sigma$ \neww{of} the ERD model (see \S~\ref{sec:ERDcomp}).
\vspace{2ex}
\end{minipage}
\end{table*}

%% file: figs_kinematics.tex
\begin{figure*}
  \includegraphics[width=2\columnwidth]{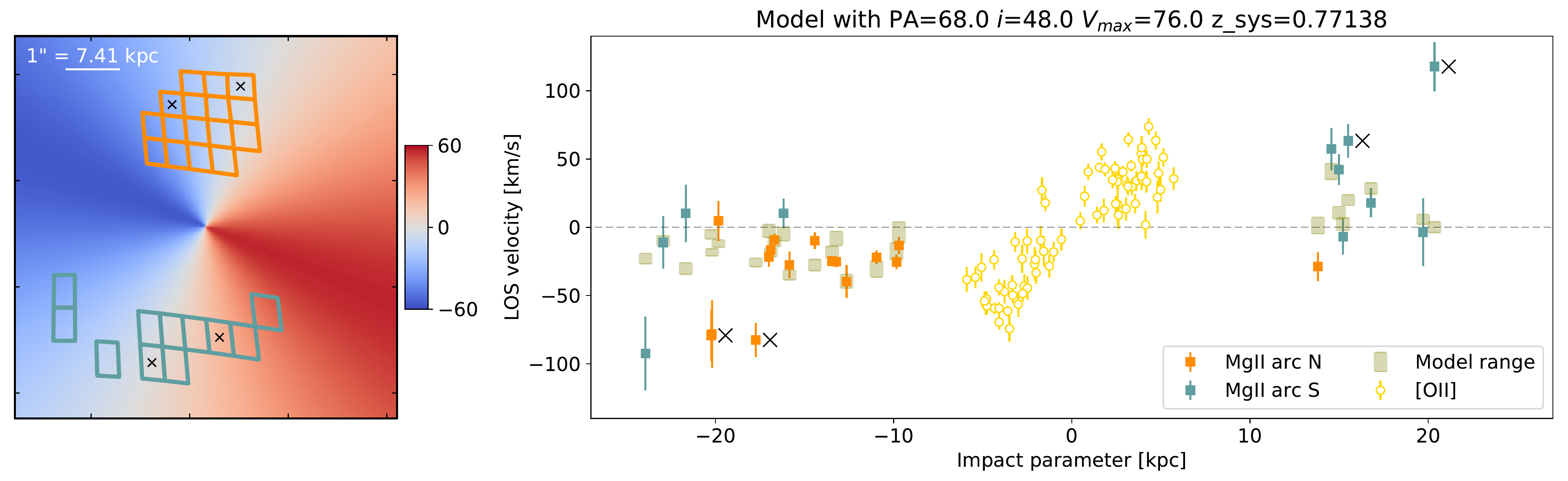} 
\caption{ 
%{Rotation curves of \mgii\ along two different orientations. Only \mgii\ detections are shown which lie inside the pseudo-slit %shown in the left panel (some of them appear on both slits). 
%{Two examples of extended rotation disk models with different position angles. 
{\it Left-hand  panel:} extended rotating-disk model (ERD) velocity \new{map} and MUSE binned spaxels (delensed) used to measure \mgii\ velocities on top of the arcs (see \S~\ref{sec:ERD}).  
{\it Right-hand panel:} \neww{Mg~{\sc ii} and [O~{\sc ii}]} velocities as a function of impact parameter. Positive (negative) impact parameters are defined to the receding (approaching) side of the model minor axis. 
Olive bars indicate the velocity interval permitted by the model within each arc spaxel (see \S~\ref{sec:ERDcomp} and \S~\ref{sec:vdisp}). \neww{Kinematic outliers (\S~\ref{sec:vdisp})} are marked with a \neww{cross} symbol \neww{in both panels}. We observe a large fraction of spaxels ($\nERD$ out of $\nhit$) being consistent (within 3$\sigma$) with the ERD model, including several seen close to the minor-axis. We emphasize that this plot {is not} a rotation curve, as our spaxels are located in a wide range of azimuthal angles with respect to the semi-major axis.} 
\label{fig:kinematics}
\end{figure*}

%% file: fig_mgii_mage_main.tex
\begin{figure}
\centering
\begin{minipage}{0.8\columnwidth}
\includegraphics[trim=0 177.5 0  -0.5, clip, width=1\textwidth]{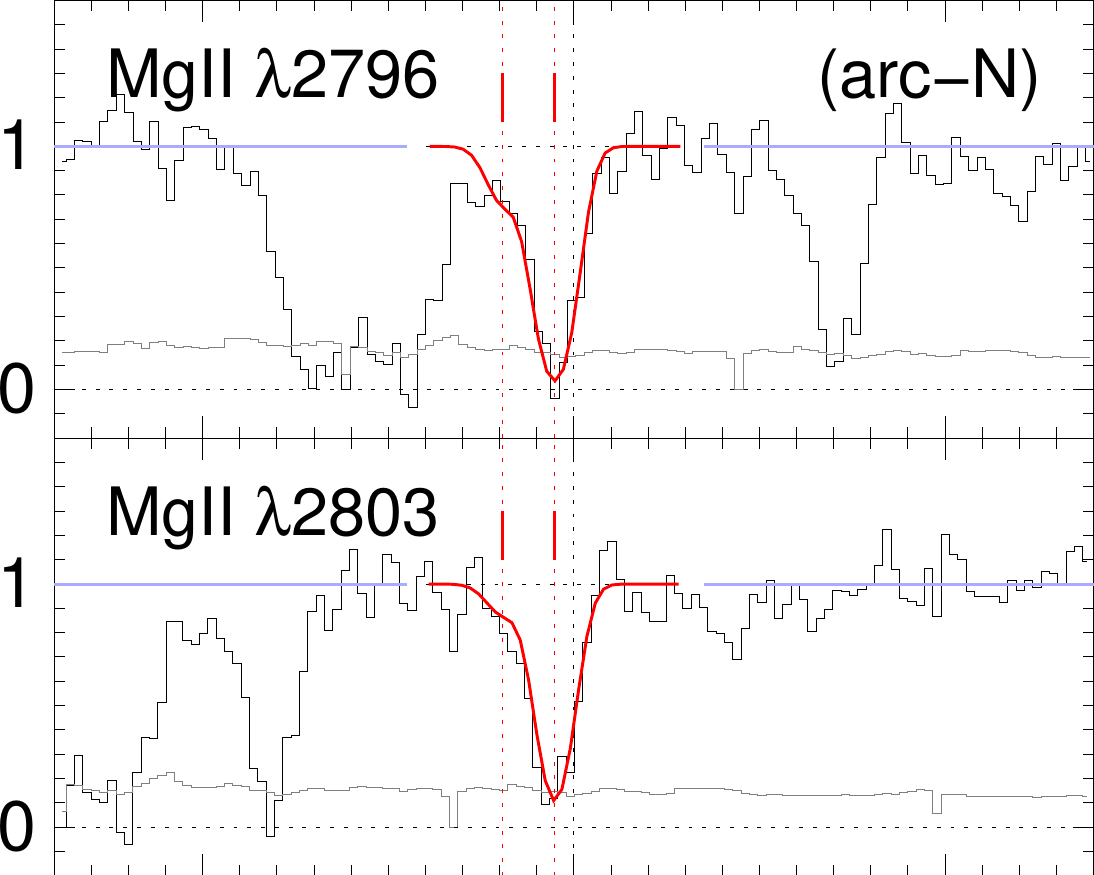}
\includegraphics[trim=0 177.5 0  -0.5, clip, width=1\textwidth]{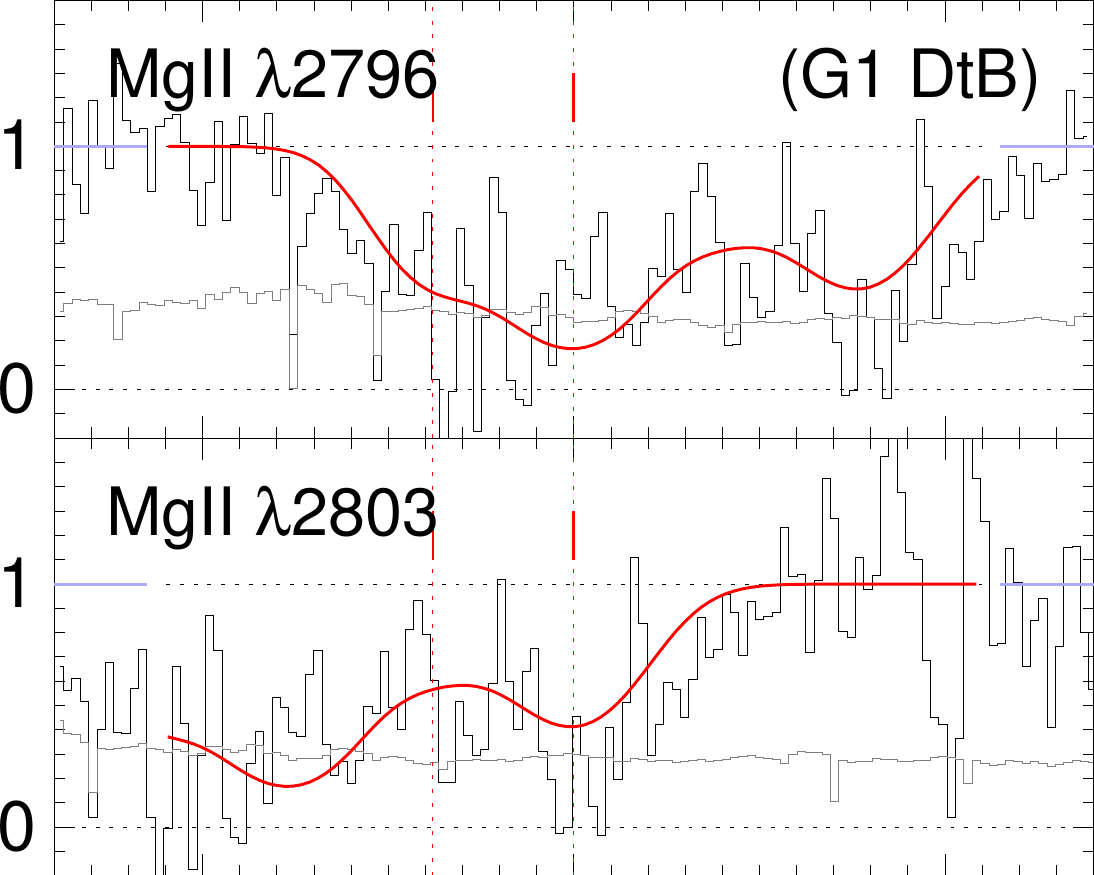}
\includegraphics[trim=0   0.0 0 380.0, clip, width=1\textwidth]{figs/objE_nor_07_Helvetica_maintext.pdf}
\end{minipage}
\caption{
{\mgii\ absorption lines observed in the Magellan/MagE spectra \neww{at $R\sim 4\,500$} on top of arc-N seen in the `East' slit (upper two panels) and
on top of G1 DtB (bottom two panels\neww{; from a special MagE `pseudo-spaxel'; see Appendix~\ref{sec:mage})}. The histograms show the normalized flux and
its associated $1\sigma$ uncertainty. The red curves are the fitted Voigt profiles. The origin of
the velocity scale is anchored to the systemic redshift $z_{\rm sys}=\zabs$ 
%established from morpho-kinematic analysis of G1's \oii\ emission 
\neww{(see} \S~\ref{sec:galpak}). We refer the reader to Appendix~\ref{sec:mage} for more details on \neww{the MagE data analysis}.}
\label{fig:mage_main}
}
\end{figure}

%% file: appendix.tex
\appendix

\section{\neww{MagE absorption-line analysis}} \label{sec:mage}

%\subsection{MagE Observations}

\neww{Here we describe the Mg~{\sc ii} absorption profile fits to the Magellan/MagE data (\S~\ref{sec:mage_main}). For the arc-N region, the typical signal-to-noise ratio ($S/N$) around Mg~{\sc ii}$\lambda\lambda 2796,2803$ at $z=\zabs$ is $S/N \approx 5$--$10$, depending on position along the slits. For the G1 DtB and arc-S regions, the $0.9\arcsec \times 1\arcsec$ binning did not provide enough $S/N$ to detect Mg~{\sc ii} in absorption. We were able to increase the $S/N$ for G1 DtB to detect Mg~{\sc ii} absorption (at the cost of enlarging the aperture) by defining a `special' pseudo-spaxel from the combination of $8$ native MagE spatial pixels ($8\times 0.3\arcsec = 2.4\arcsec$) centred around G1's [O~{\sc ii}] emission, from both MagE slits (E and W).
}

\input{fig_mgii_mage}

\mgii\ is detected to sensitive limits at four positions of the pseudo-spaxels of the MagE W slit (\#7, \#8, \#9, and \#10) and at two pseudo-spaxels of
the MagE E slit (\#10 and \#11; see Fig.~\ref{fig:mage_slits}). All these detections correspond to the arc-N region where the $S/N$ is larger, and the spectra are shown in Fig.~\ref{fig:magefits}. We note that given the seeing of the observations we detect continuum emission on adjacent spaxels that would otherwise not show light based on the \HST\ image. With the exception of G1 DtB (see below), \mgii\ is not confidently detected at other positions due
to low $S/N$. These include the portions of arc-S covered by both MagE slits. \mgi\ is not detected at all and possible \feii\ lines \neww{fall} into the Lyman-$\alpha$ forest \neww{and were not studied}.

The resolving power of MagE is twice that of MUSE. This advantage reveals an edge-leading asymmetry in the arc-N \mgii\ absorption profiles (see \S~\ref{sec:corotating}), which is most clearly seen at W slit pseudo-spaxels \#8
and \#10, where the $S/N$ is highest. We used the \vpfit\ software \citep{Carswell2014} to perform Voigt-profile fitting using two components
located at different redshifts and \neww{having} independent column densities but equal velocity broadening \neww{to} minimize
the number of free parameters and provide
robust constraints on redshifts. The fits are shown by the red lines in Fig.~\ref{fig:magefits}. The reduced $\chi^2$ values of our fits are in the range [0.6,1.5].

Details on the fits are given in Table~\ref{table:mgii_mage}. Since the lines are saturated, column densities cannot be reliably determined and so we do not report them. Instead, we use the obtained values of column density and Doppler parameters to infer the equivalent widths of the fitting profiles using the approximation given in \citet{Draine2011}.

The Doppler parameters derived from both components are
usually small ($b\sim 30$\,\kms) \neww{comparable} to the instrumental resolution of MagE
($b_{\rm inst}\approx 28$\,\kms). The mean velocity offset between the two components
is found to be $\approx 117$\,\kms. The strongest component is the reddest and closest in velocity
to the galaxy's systemic redshift inferred from the \oii\ emission.

As mentioned above, \mgii\ absorption is also detected on top of G1 DtB, from the combination of the `special' pseudo-spaxel that combines data from both slits (E and W; see above). As seen in Fig.~\ref{fig:magefits}, the
absorption there is particularly strong producing a trough where the two \mgii\ transition lines are
blended together (consistent with what we observe with MUSE; see \S~\ref{sec:abs}). Since the S/N is low, we can only decompose the profile into two broad components located at about the systemic redshift and $-380$\,\kms, respectively.

\input{table_mgii_mage}
\input{tab_transitions_air}

\section{G1 measurements}
Table~\ref{table:g1_hst} presents the \hst\ photometry of G1 corrected by Galactic extinction, but in the image plane (i.e., magnified).

\new{Figure~\ref{fig:ppxf} presents the de-magnified integrated MUSE spectrum of G1, together with the \ppxf\ and emission line fit.}

\input{tab_hst}
\input{fig_ppfx}

%% file: fig_mgii_mage.tex
\begin{figure}
\centering
\includegraphics[width=0.9\columnwidth]{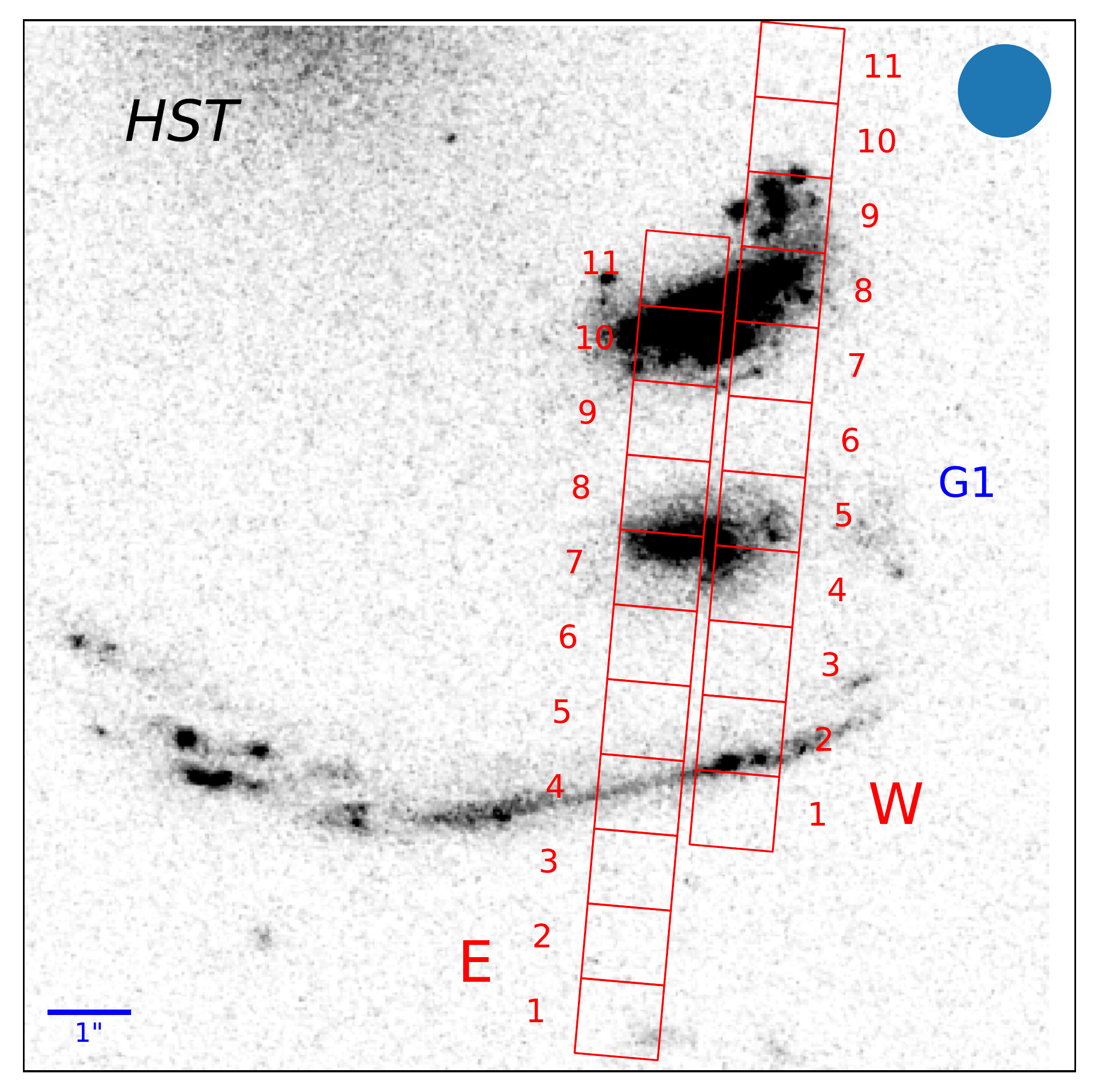}
\caption{Location of the MagE slits in the image plane, from left to right: East (`E') and West (`W'), respectively. Pseudo-spaxels are labeled from $\#1$ to $\#11$ in each slit from South to North. The background image corresponds to \HST/F814W (as in Fig.~\ref{fig:fov}), \neww{highlighting} the spatial structure within each pseudo-spaxel; however, \neww{we note that this signal is} effectively smeared out by the seeing (indicated by the blue circle).}

%The background image corresponds to \HST/F814W as in Figure~\ref{fig:fov}. 
%The effective seeing PSF of the observations is indicated by the blue circle. 
%We note that with this PSF we detect continuum emission on adjacent spaxels that would normally not show light based on the \HST\ image. \CL{I have modified this sentence to improve clarity.} NT: moved this text to the appendix main text.

\label{fig:mage_slits}

\end{figure}

\begin{figure*}
\includegraphics[width=0.5\columnwidth]{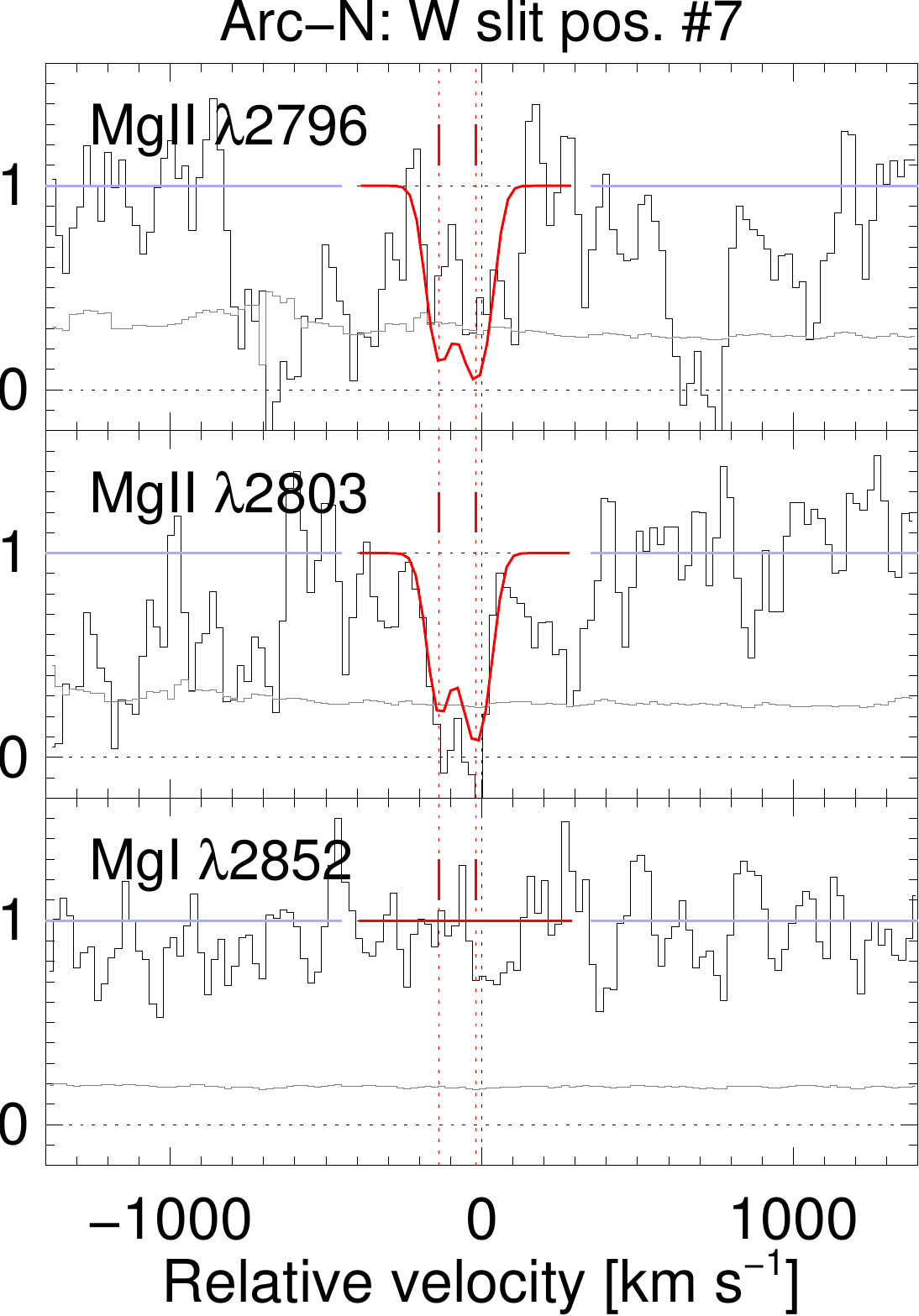}
\includegraphics[width=0.5\columnwidth]{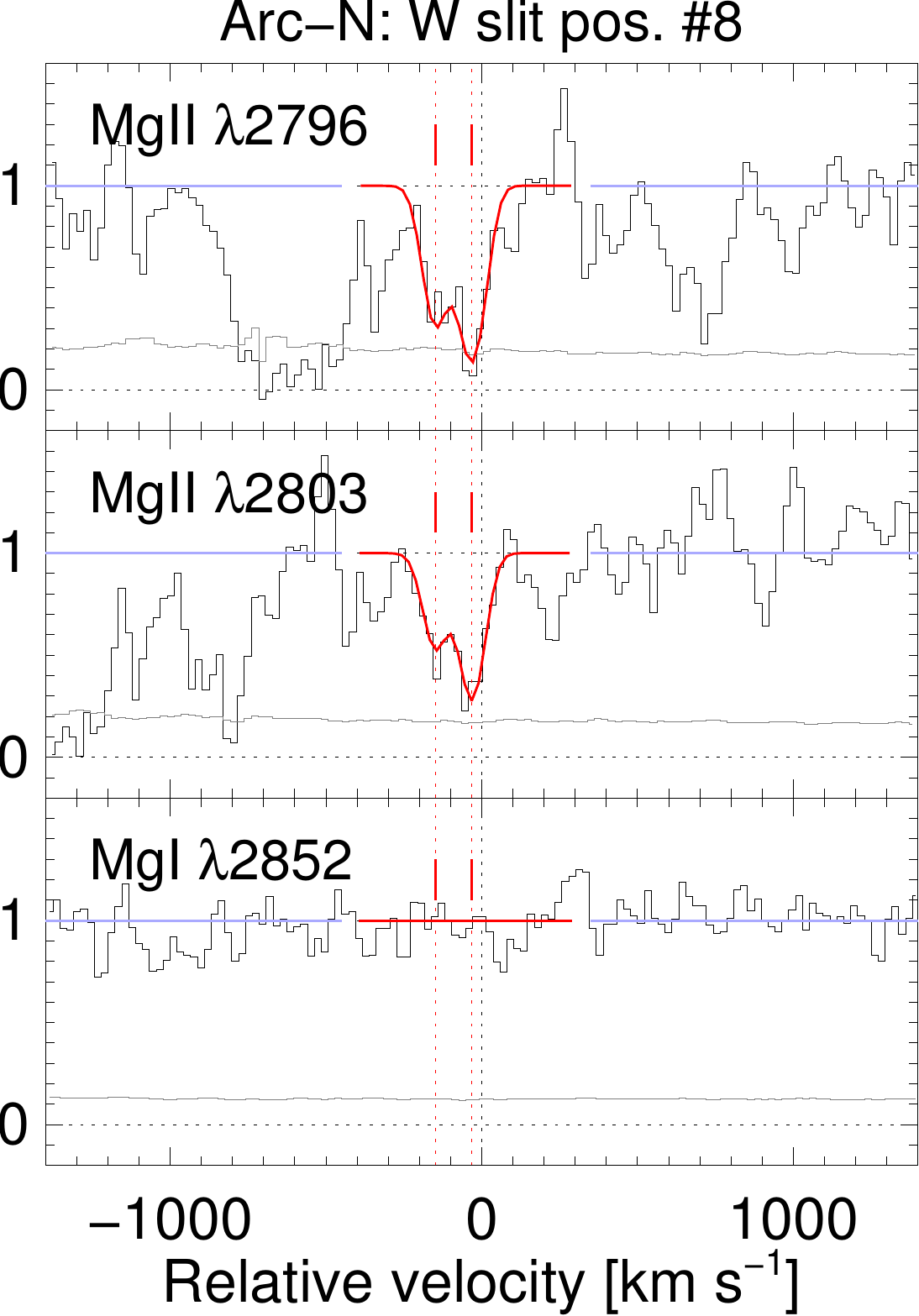}
\includegraphics[width=0.5\columnwidth]{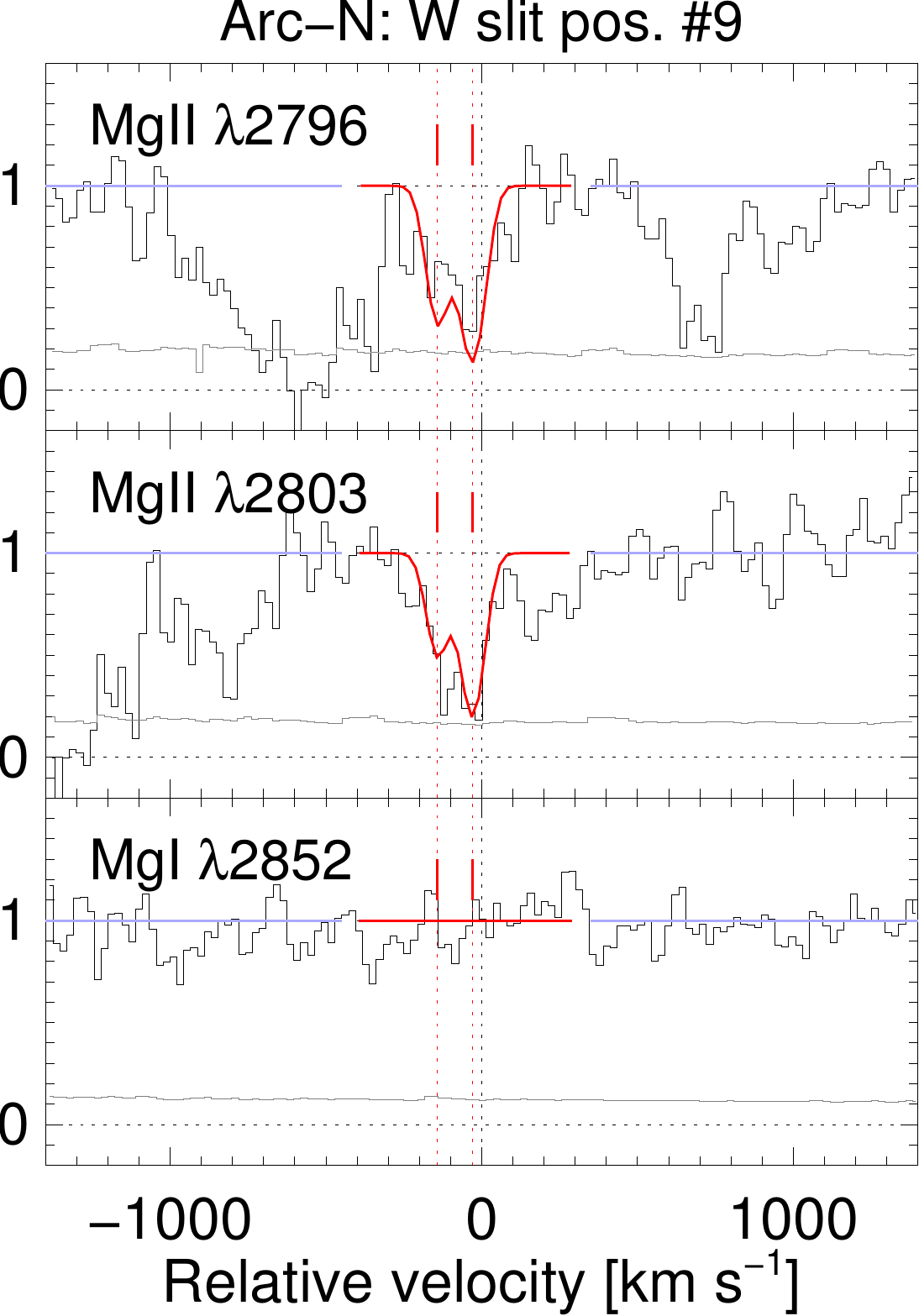}
\includegraphics[width=0.5\columnwidth]{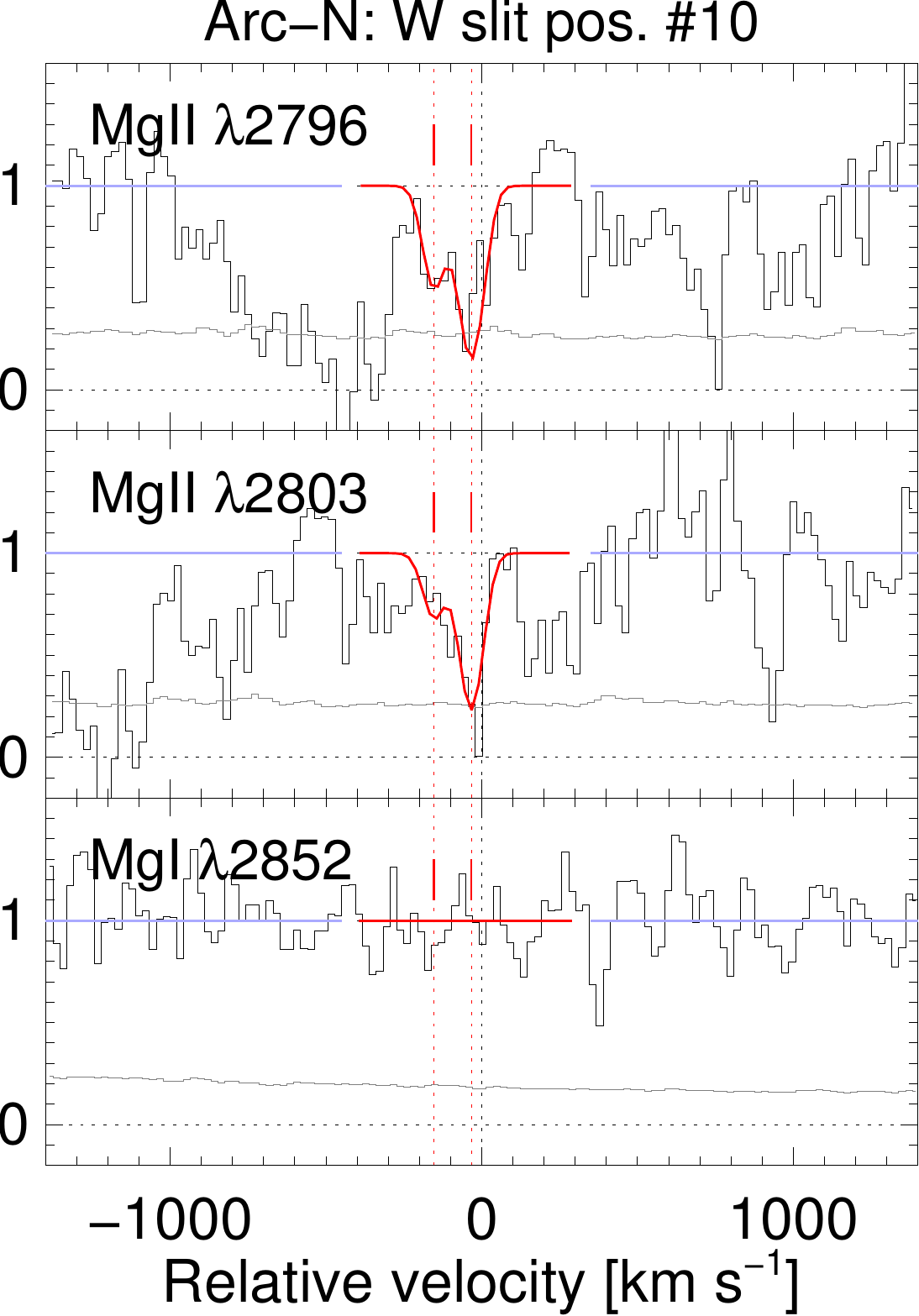}\\
\vspace{0.5cm}
\includegraphics[width=0.5\columnwidth]{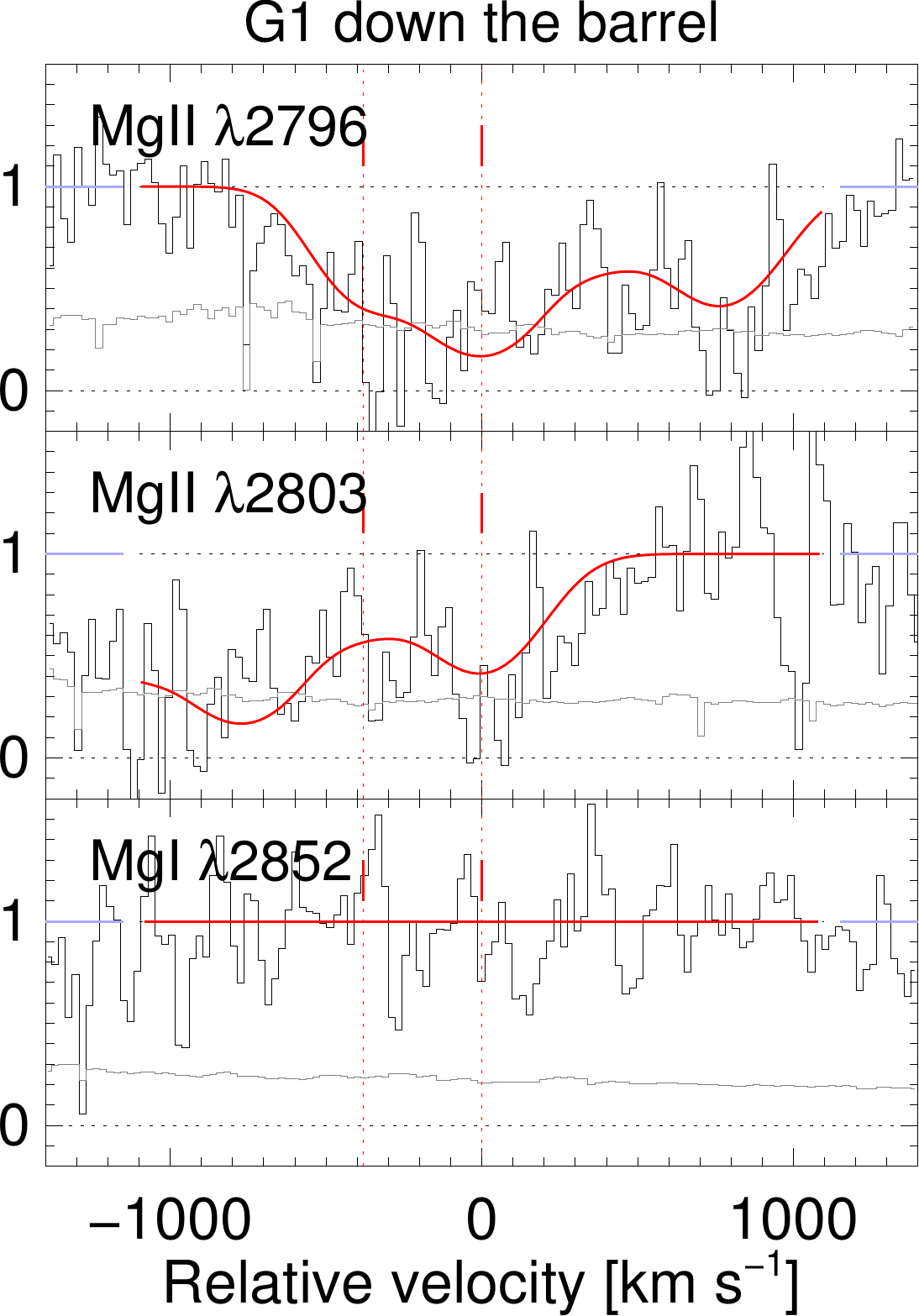}
\phantom{\includegraphics[width=0.5\columnwidth]{figs/objE_nor_07_Helvetica.pdf}}
\includegraphics[width=0.5\columnwidth]{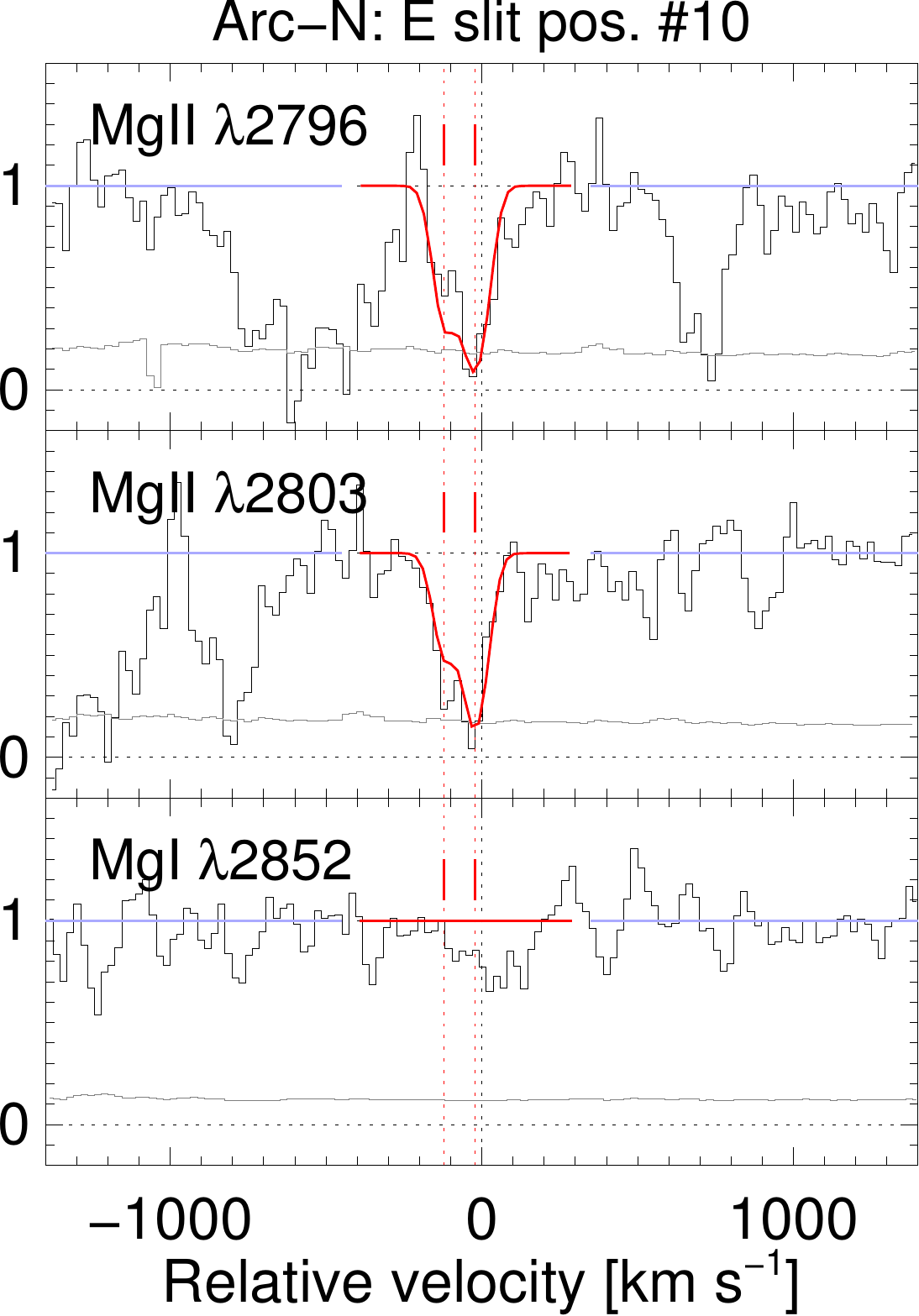}
\includegraphics[width=0.5\columnwidth]{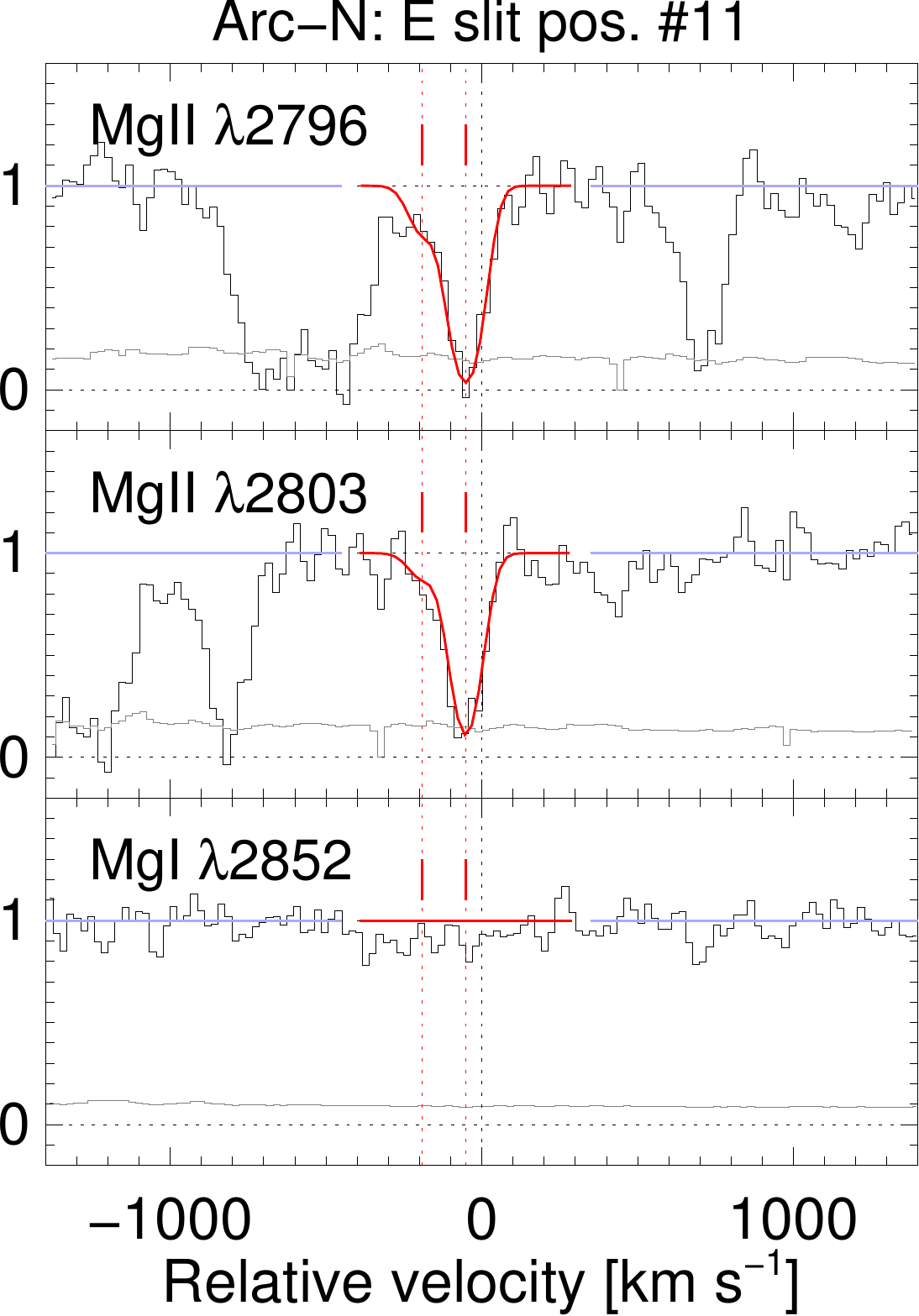}
\caption{
{\mgii\ absorption lines detected in MagE pseudo-spaxels W\#7 to
\#10 on top of arc-N (upper panels), in pseudo-spaxels E\#10 and \#11 also on top of arc-N (bottom-right panels), and in the special
pseudo-spaxel on top of G1's DtB region (bottom-left panel). The histograms show the normalized flux and its associated $1\sigma$ uncertainty.
The red curves are the fitted Voigt profiles. The expected location of \ion{Mg}{i}\,$\lambda$2852 which is displayed in each panel illustrates
the typical noise of the data but since the line is not detected no fit was performed. The origin of the velocity scale is anchored to the systemic
redshift $z=\zabs$ established from the morpho-kinematic analysis of G1's \oii\ emission (see \S~\ref{sec:galpak}).
}
\label{fig:magefits}
}
\end{figure*}

%% file: table_mgii_mage.tex
\begin{table*}
\caption{Absorption-line properties from MagE data.}
\label{table:mgii_mage}
\centering
\begin{tabular}{cccccc}
\hline\hline
Region & MagE     & $v({\rm comp}_2)$ & $v({\rm comp}_1)$ & $b$    & W$_{0}($2796)\\
       &  Spaxel  &    (\kms)         & (\kms)            & (\kms) & (\AA)        \\
(1)    & (2)      & (3)     & (4)   & (5)    & (6)              \\
\hline
arc-N & W\#7  &  $-136.4\pm 14.7$  & $-18.1\pm 12.6$ & $\lesssim 25$\,$^{\rm a}$ & $1.90 \pm 0.35 $\\
arc-N & W\#8  &  $-149.2\pm  6.6$  & $-31.3\pm  4.7$ & $32.7\pm  5.8$        & $1.55 \pm 0.24$ \\
arc-N & W\#9  &  $-142.0\pm  7.9$  & $-29.8\pm  5.8$ & $\lesssim 25$\,$^{\rm a}$ & $1.41 \pm 0.13 $\\
arc-N & W\#10 &  $-153.1\pm 11.9$  & $-33.8\pm  6.4$ & $25.1\pm 11.5$       &  $1.24 \pm 0.56 $ \\
arc-N & E\#10 &  $-120.8\pm 12.8$  & $-20.9\pm  8.5$ & $\lesssim 25$\,$^{\rm a}$ & $1.48 \pm 0.20$ \\
arc-N & E\#11 &  $-191.8\pm 28.7$  & $-51.4\pm  4.3$ & $44.6\pm  7.4$        & $1.53 \pm 0.28 $ \\
DtB & \ special\,$^{\rm b}$  &  $\sim -380$  &  $\sim 0$  &  $\sim 210$  & \ \ $6.45 \pm 0.70 $\,$^{\rm c}$ \\

\hline
\end{tabular}\\
\vspace{1ex}
\begin{minipage}{0.6\textwidth}
\raggedright {\bf Notes:} 
(1)~Region probed at this slit position;
(2)~MagE slit and spaxel position number (see Figure~\ref{fig:mage_slits});
%(3)~Impact parameter of the spaxel to G1 in the absorber plane; positive values are arbitrarily defined to be on the receding side of the velocity model, and negative values otherwise (see \S~\ref{sec:results});
(3)~and (4) Rest-frame velocity of each component with respect to the
systemic redshift $z=\zabs$;
(5)~Doppler parameter of each \neww{individual} component;
(6)~Rest-frame equivalent width of the entire \ion{Mg}{ii}\,$\lambda$2796
absorption (i.e., modeled summing up the contributions of both components from the fits).\\
$^{\rm a}$: unresolved components.\\
$^{\rm b}$: special pseudo-spaxel that combines eight native MagE pixels ($8\times 0.3\arcsec = 2.4\arcsec$) centred around G1's \oii\ emission, from both MagE slits (E and W).\\
$^{\rm c}$: may be overestimated due to continuum placement uncertainties in the low S/N spectrum.
\vspace{2ex}

\end{minipage}
\end{table*}

%% file: tab_transitions_air.tex
\begin{table}
\caption{Rest-frame transition wavelengths.}
\label{table:wv_air}
% \centering
\begin{tabular}{l c c}
\hline\hline
Transition & Wavelength  \\
 & in air (\AA)$^{\rm a}$ \\
(1) & (2) \\
\hline
Mg~{\sc ii} 2796  &  2795.528 \\ 
Mg~{\sc ii} 2803  &  2802.705 \\
$[$O~{\sc ii}$]$ 3727  &  3726.032 \\
$[$O~{\sc ii}$]$ 3729  &  3728.815 \\
\hline
\end{tabular}\\ 
\vspace{1ex}

\begin{minipage}{0.3\textwidth}
\raggedright {\bf Notes:}\\ 
$^{\rm a}$: From \url{https://physics.nist.gov}
\vspace{2ex}
\end{minipage}

\end{table}

%% file: tab_hst.tex
\begin{table}
\caption{G1 photometry.}
\label{table:g1_hst}
% \centering
\begin{tabular}{l c}
\hline\hline
Filter & Magnitude\,$^{\rm a,b}$  \\
       & (AB)       \\
(1)    & (2)        \\ 
\hline
ACS/F606W & $22.50 \pm 0.10 $ \\ 
ACS/F814W  & $21.46 \pm 0.08 $ \\
WFC3/F110W  & $20.87 \pm 0.07 $ \\
WFC3/F160W  & $20.57 \pm 0.07 $ \\
\hline
\end{tabular}\\
\vspace{1ex}
\begin{minipage}{0.3\textwidth}
\raggedright {\bf Notes:}\\ 
$^{\rm a}$: Corrected by Galactic extinction using \citet{extinction}. \\
$^{\rm b}$: Magnified values.
\vspace{2ex}
\end{minipage}
\end{table}

%% file: fig_ppfx.tex
\begin{figure*}
\begin{minipage}{1\textwidth}
%\flushright
   \includegraphics[width=1.0\textwidth]{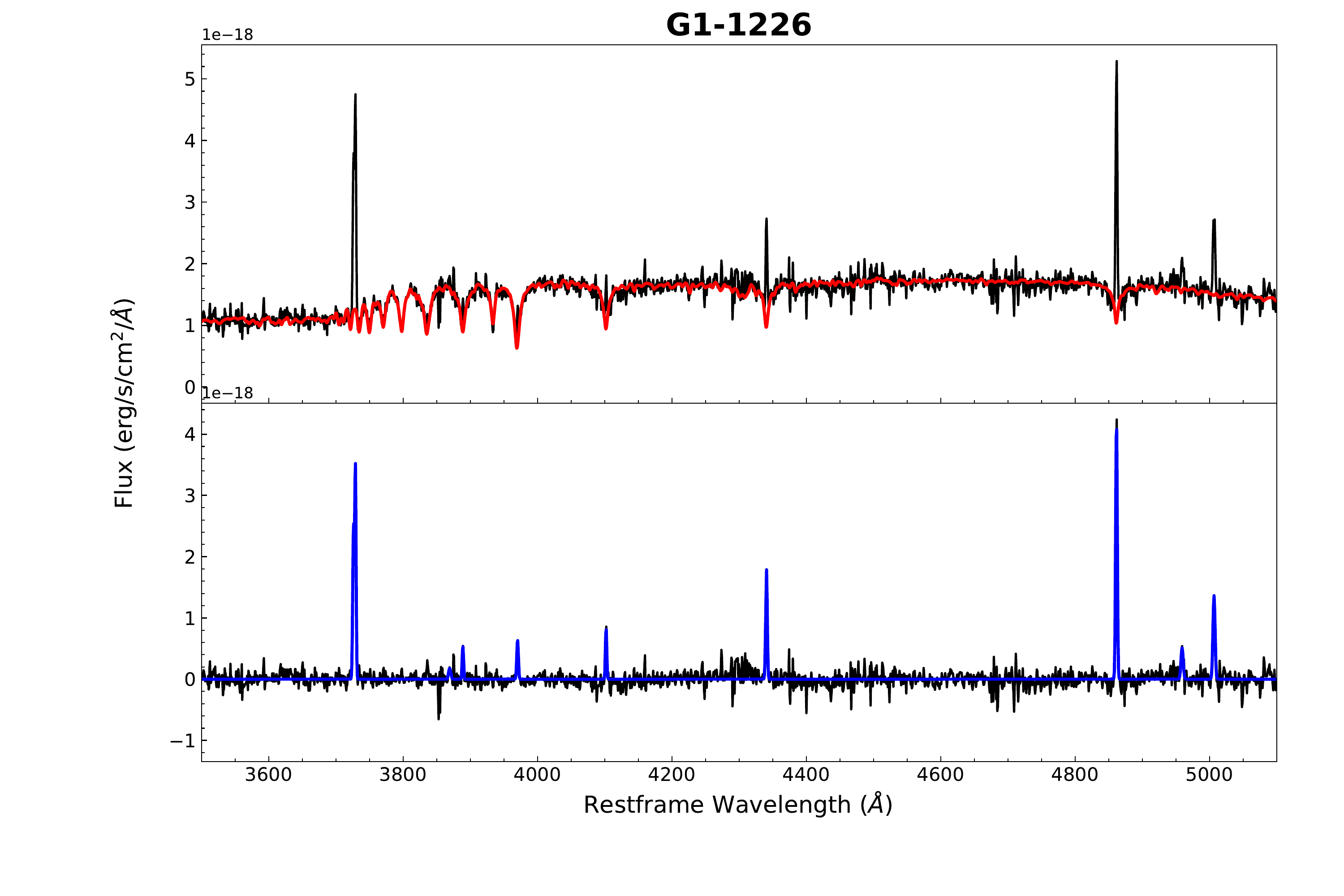}
\end{minipage}
\caption{\new{{\it Top-panel:} de-magnified observed (black) rest-frame optical spectrum of G1 extracted from the MUSE datacube together with the stellar light model by \ppxf\ (red) after masking out emission lines. {\it Bottom-panel:} Residual spectrum after subtracting the \ppxf\ stellar light model; we used this residual spectrum to fit emission lines (blue) from which we obtained emission line fluxes.}}
\label{fig:ppxf}
\end{figure*}